%% file: main-asplos.tex
\documentclass[conference]{IEEEtran}

\input{commands}
\usepackage{filecontents}
\usepackage{color,colortbl,xcolor}
\usepackage{tcolorbox}
\usepackage{listings}
\usepackage[labelfont=bf]{caption}
\usepackage{subcaption}
\usepackage{tikz}
\usepackage{threeparttable} 
\usepackage{soul} 
\usetikzlibrary{pgfplots.groupplots}

\definecolor{dkgreen}{rgb}{0,0.6,0}
\definecolor{gray}{rgb}{0.5,0.5,0.5}
\definecolor{mauve}{rgb}{0.58,0,0.82}
\definecolor{orangep}{rgb}{0.71, 0.43, 0.89}
\definecolor{orp}{rgb}{1, 0.7, 0.278}

\definecolor{darkBlue}{rgb}{0.000000,0.000000,0.545098}
\definecolor{darkGreen}{rgb}{0.000000,0.392157,0.000000}
\definecolor{DarkGray}{gray}{0.4}
\definecolor{javared}{rgb}{0.6,0,0} 
\definecolor{javagreen}{rgb}{0.25,0.5,0.35} 
\definecolor{javapurple}{rgb}{0.5,0,0.35} 
\definecolor{javadocblue}{rgb}{0.25,0.35,0.75} 
\definecolor{lightgray}{gray}{0.95}
\definecolor{shadecolor}{RGB}{150,150,150}
\definecolor{blueA}{RGB}{204,229,255}
\definecolor{redA}{RGB}{112,0, 0}
\lstdefinestyle{MyJavaSmallStyle} {
  language=C++,
  frame=none,
  xleftmargin=15pt,
  stepnumber=1, 
  numbers=left, 
  numbersep=5pt,
  numberstyle=\tiny\color[gray]{0.177}, 
  belowcaptionskip=\bigskipamount,
  captionpos=b, 
  escapeinside={*'}{'*},
  tabsize=5,
  emphstyle={\bf},
  escapechar=!,
  basicstyle=\scriptsize\ttfamily,
  keywordstyle=\color{javapurple}\bfseries,
  stringstyle=\color{javared},
  commentstyle=\color{javagreen},
  morecomment=[s][\color{javadocblue}]{/**}{*/},
  showspaces=false,
  columns=flexible,
  showstringspaces=false,
  morecomment=[l]{//},
  tabsize=2,
  breaklines=true,
  moredelim=[is][\underbar]{^}{^}
}

\newcolumntype{R}[1]{>{\raggedleft\arraybackslash}p{#1}}
\newcolumntype{L}{>{\raggedright\arraybackslash}c}
\usepackage{booktabs}

\newcommand{\tool}{{\sc PerfGen}\xspace}
\newcommand{\perfgen}{{\sc PerfGen}\xspace}

\newcommand{\jason} [1] {\textcolor{teal}{{Jason}: #1}}
\newcommand{\todo} [1] {\textcolor{red}{{TODO}: #1}}

\def \codefont#1{\texttt{#1}}

\newcommand{\EvalResultSpeedupX}{43X\xspace}
\newcommand{\EvalResultIterPct}{0.004\%\xspace}
\newcommand{\EvalResultPhIterPct}{4.94\%\xspace}
\newcommand{\EvalResultPhTimePct}{86.28\%\xspace}
\newcommand{\EvalResultMutationSpeedup}{1.81X\xspace}
\newcommand{\CollatzProgramName}{\textit{Collatz}\xspace}

\soulregister{\tool}{7}
\soulregister{\todo}{7}
\soulregister{\jason}{7}
\soulregister{\codefont}{7}
\soulregister{\EvalResultSpeedupX}{7}
\soulregister{\EvalResultIterPct}{7}
\soulregister{\EvalResultPhIterPct}{7}
\soulregister{\EvalResultPhTimePct}{7}
\soulregister{\EvalResultMutationSpeedup}{7}
\soulregister{\CollatzProgramName}{7}

\usepackage[normalem]{ulem}

\begin{document}

\title{
PerfGen: Automated Performance Benchmark Generation for Big Data Analytics}

\date{}
\author{
\IEEEauthorblockN{
Jiyuan Wang\IEEEauthorrefmark{1}, 
Jason Teoh\IEEEauthorrefmark{1}, 
Muhammand Ali Gulza\IEEEauthorrefmark{2}, 
Qian Zhang\IEEEauthorrefmark{3}, 
Miryung Kim\IEEEauthorrefmark{1}
}
\IEEEauthorblockA{
\IEEEauthorrefmark{1}\textit{University of California, Los Angeles}, USA \\
\{wangjiyuan, jteoh, miryung\}@cs.ucla.edu}
\IEEEauthorblockA{
\IEEEauthorrefmark{2}\textit{Virginia Tech}, USA \\
magulza@vt.edu}
\IEEEauthorblockA{
\IEEEauthorrefmark{3}\textit{University of California, Riverside}, USA \\
qzhang@cs.ucr.edu}
}

\maketitle

\thispagestyle{empty}

\begin{abstract}
Many symptoms of poor performance in big data analytics such as {\em computational skews}, {\em data skews}, and {\em memory skews} are input dependent. However, due to lack of inputs that can trigger such performance symptoms, it is hard to debug and test big data analytics.

We design \tool to automatically generate inputs for the purpose of performance testing.  \tool overcomes three challenges when naively using automated fuzz testing for the purpose of performance testing. First, typical greybox fuzzing relies on coverage as a guidance signal and thus is unlikely to trigger interesting performance behavior. Therefore, \tool provides performance monitor templates that a user can extend to serve as a set of guidance metrics for grey-box fuzzing. Second, performance symptoms may occur at an intermediate or later stage of a big data analytics pipeline. Thus, \tool uses a {\em phased fuzzing} approach. This approach identifies symptom-causing intermediate inputs at an intermediate stage first and then converts them to the inputs at the beginning of the program with a pseudo-inverse function generated by a large language model. Third, \tool defines sets of skew-inspired input mutations, which increases the chance of inducing performance problems. We evaluate \tool using four case studies. \tool achieves at least \EvalResultSpeedupX speedup compared to a traditional fuzzing approach when generating inputs to trigger performance symptoms. Additionally, identifying intermediate inputs first and then converting them to original inputs by pseudo-inverse functions, which only takes up to 2 prompting iterations using a large language model, enables \tool to generate such workloads in less than \EvalResultIterPct of the iterations required by a baseline approach. 

\end{abstract}
\maketitle
\input{introduction}
\input{motivating_collatz}

\input{approach}

\input{evaluations_2022}

\input{related}
\input{conclusion}

\bibliographystyle{plain}
\bibliography {bibs/thesis-merged}

\end{document}

%% file: commands.tex
\newcommand{\eat}[1]{}

\usepackage{latexsym}
\usepackage{amsfonts}
\usepackage{amsmath}
\usepackage{amssymb}
\usepackage{color}
\usepackage{colortbl}
\usepackage{epsfig}
\usepackage{xspace}
\usepackage{graphicx}
\usepackage{enumerate}
\usepackage{comment}
\usepackage{stmaryrd}
\usepackage[all]{xy}

\usepackage{algorithm}
\usepackage{epsfig}
\usepackage{multirow}
\usepackage{hhline}
\usepackage{url}

\usepackage{breakurl}

\newcommand{\at}[1]{\protect\ensuremath{\mathsf{#1}}\xspace}

\newcommand{\bi}{\begin{itemize}}
\newcommand{\ei}{\end{itemize}}

\newcommand{\be}{\begin{enumerate}}
\newcommand{\ee}{\end{enumerate}}
\newcommand{\beqn}{\begin{eqnarray*}}
\newcommand{\eeqn}{\end{eqnarray*}}

\newcommand{\ie}{\emph{i.e.,}\xspace}
\newcommand{\eg}{\emph{e.g.,}\xspace}

\newcounter{ccc}

\newcommand{\eop}{\hspace*{\fill}\mbox{$\Box$}\vspace{1ex}}     
\newcounter{example}

\newcommand{\nthesection}{\arabic{section}}
\newcounter{prop}[section]

\newcounter{lemma}[section]

\newcounter{cor}
\renewcommand{\thecor}{\arabic{theorem}}

\newcounter{definition}[section]

\newcounter{alg}[section]
\renewcommand{\thealg}{\nthesection.\arabic{alg}}

\newcounter{arule}
\renewcommand{\thearule}{\arabic{arule}}

\newcounter{claim}
\renewcommand{\theclaim}{\arabic{claim}}

\DeclareMathAlphabet\mathbfcal{OMS}{cmsy}{b}{n}
\DeclareMathAlphabet{\mathbfsf}{\encodingdefault}{\sfdefault}{bx}{n}

\usepackage{amsmath}
\usepackage{listings}
\usepackage{color}
\usepackage[normalem]{ulem}
\usepackage{amssymb}
\usepackage{tablefootnote}
\usepackage{pifont}
\makeatletter
\newcommand*{\rom}[1]{\expandafter\@slowromancap\romannumeral #1@}
\makeatother

\makeatletter
\let\oldnl\nl
\newcommand{\nonl}{\renewcommand{\nl}{\let\nl\oldnl}}
\makeatother

\usepackage[noend]{algpseudocode}

\usepackage{times}
\usepackage{array}
\usepackage{url}
\usepackage{pbox}
\usepackage{float}
\usepackage{subcaption}
\usepackage{pgfplots}
\usepackage{filecontents}
\usepackage{balance}
\usepackage[labelfont=bf]{caption}
\usepackage{makecell}
\usetikzlibrary{patterns}
\usetikzlibrary{intersections}
\usetikzlibrary{calc}
\usetikzlibrary{arrows.meta}
\usetikzlibrary{shapes.misc}

\newcolumntype{L}{>{\centering\arraybackslash}m{7cm}}
\newcolumntype{K}{>{\centering\arraybackslash}m{3.7cm}}

\definecolor{orangep}{rgb}{0.71, 0.43, 0.89}
\definecolor{orp}{rgb}{1, 0.5, 0.2}
\definecolor{dkgreen}{rgb}{0,0.6,0}
\definecolor{gray}{rgb}{0.3,0.3,0.3}
\definecolor{mauve}{rgb}{0.58,0,0.82}

\lstdefinelanguage{Scala}{
  keywords={typeof, class, override, new, true, false, catch,def,val, function, return, null, catch, switch, var, private, if, in, while, do, else, case, break, trait, extends},
  keywordstyle=\color{mauve}\bfseries,
  ndkeywords={String,ProvenanceRDD,RDD,Float, Boolean, TaintedString, TaintedFloat, SingleOrderedTracker, ProvenanceRow,InfluenceTracker,OrderingInfluenceTracker ,TaintedInt, TaintedAny, Ordering, Provenance, FilterInfluenceFunction, InfluenceFunction},
 ndkeywordstyle=\color{orange}\bfseries,
  otherkeywords={+, =>,<=, ==, >,< , ||},
  identifierstyle=\color{black},
  sensitive=false,
  comment=[l]{//},
  morecomment=[s]{/*}{*/},
  commentstyle=\color{purple}\ttfamily,
  stringstyle=\color{red}\ttfamily,
  morestring=[b]',
  morestring=[b]",
  moredelim=**[is][\color{red}]{@}{@},
}
\lstdefinestyle{fault}{ numbers=none, xleftmargin=1.5em , otherkeywords={ =>,<=, ==, > , ||}}
\include{data}  

\usepackage{enumitem}
\setitemize{itemsep=4pt,topsep=4pt,parsep=0pt,partopsep=0pt}


%% file: introduction.tex
\vspace{-1ex}
\section{Introduction}
\vspace{-1ex}
As the capacity to store and process data has increased remarkably, large scale data processing has become an essential part of software development. Data-intensive scalable computing (DISC) systems, such as MapReduce~\cite{Dean2008:MapReduce}, 
Apache Hadoop~\cite{Hadoop}, and Apache Spark~\cite{spark}, have shown great promises to address the \textit{scalability} challenge.
These systems abstract away the complex execution models, deep software stacks, and numerous customizable configurations from users. However, this comes at the cost of {\em performance incomprehensibility}, \ie novice developers lack the necessary knowledge to prevent and correct performance issues. This problem is exacerbated when {\em performance variability is input dependent} and {\em existing test data fails to expose pathological performance symptoms}. 

\begin{figure} 
 \centering
 \includegraphics[width=0.9\linewidth]{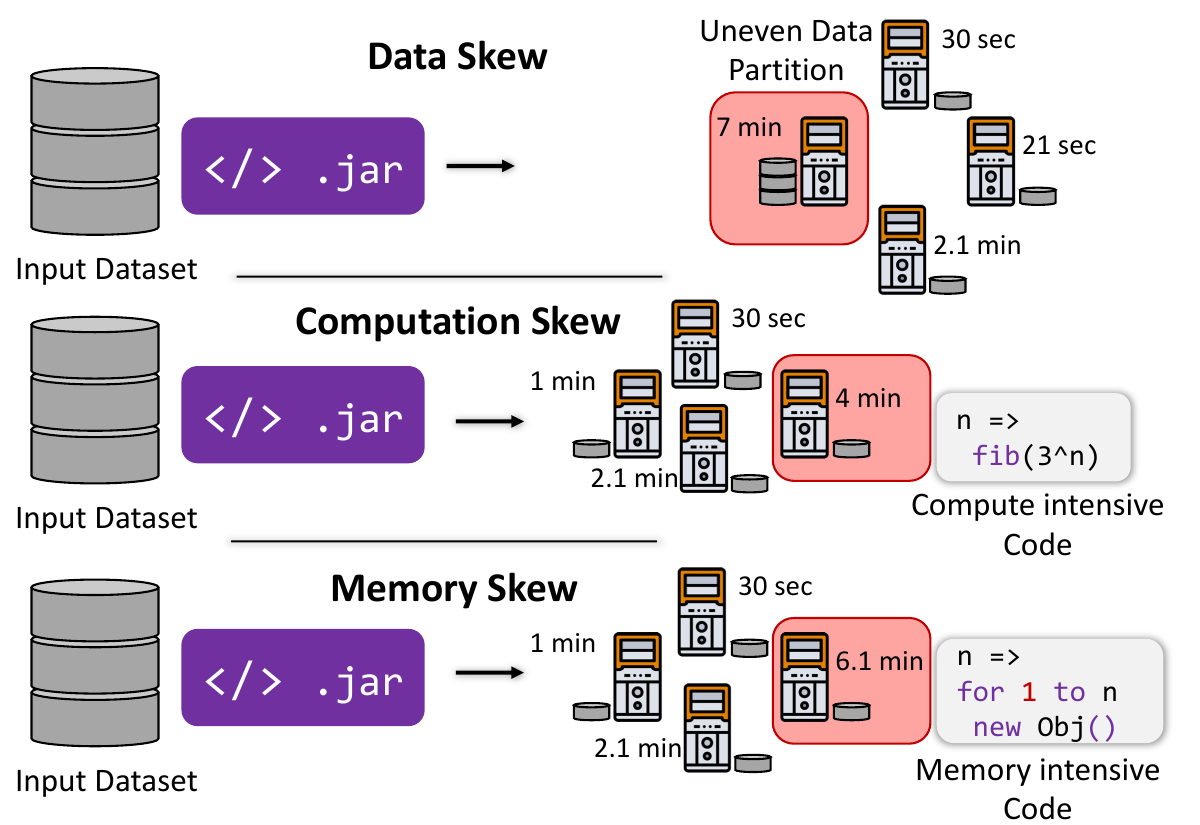}
 \vspace{-2ex}
 \caption{Three sources of performance skews}
 \label{fig:skew}
 \vspace{-5ex}
\end{figure} 
Prior work~\cite{teoh:socc:perfdebug} has discussed several major sources of performance slowdowns in DISC applications. Figure~\ref{fig:skew} visualizes three kinds of performance problems. \emph{Data skew}~\cite{Kwon:SkewTune} happens when the input is unevenly distributed across computation nodes, leading to a few nodes taking a much bigger workload and becoming stragglers. \emph{Computational skew}~\cite{teoh:socc:perfdebug} occurs when a normal input triggers an extremely long, intensive computation (\eg finding the n-$th$ Fibonacci number). \emph{Memory Skew}~\cite{blackburn:2004:sigm} occurs when a certain record triggers repetitive, memory-intensive operations (\eg \texttt{malloc} or object creation). Performance skews are often {\em input dependent}. Therefore, our goal is to automatically generate test inputs that can trigger performance skews. Such inputs could guide developers in correcting the root cause of performance bottlenecks.

\noindent \textbf{Challenges.} The standard practice for testing data-intensive applications today is to select a subset of inputs based on the developers' hunch with the hope that it will reveal possible performance issues. For example, to test big data applications, developers may use a small sample of data selected via random sampling or top $k$ sampling. Not surprisingly, such sampling is unlikely to yield performance skews described above.  Developers could always increase the number of samples, or modify the samples in different ways; however, such adhoc test input generation will significantly increase testing time.

Fuzz testing has been proven to be highly effective in revealing a diverse set of bugs, including performance defects~\cite{memlock:2020:fuzzing, lemieux2018perffuzz, slowfuzz:2017:ccs, badger:2018:issta},
correctness bugs~\cite{JQF,zest, fudge:google:2019}, and security vulnerabilities~\cite{tegan:2020:icse, collafl:2018:sp, ruiter:2015:usenix, juraj:2016:ccs}. 
Generally speaking, fuzzing techniques start from a seed input, run the program on the selected input, generate new inputs by mutating the previous input, and add new inputs to the queue if they improve a given guidance metric such as branch coverage. However, it is nontrivial to apply such traditional fuzzing to data-intensive applications for the following reasons. 

First, the {\em long latency} of big data analytics prohibits repetitively invoking  the entire application from the beginning\textemdash a common assumption for fuzz testing techniques. Inherently, these are two conflicting goals: eliminating long-latency during testing vs.~encountering performance skews that are rare during testing. In particular, when the performance skew symptoms appear at a later computation stage, it is difficult to generate inputs in the beginning of the program, which induce performance problems in the later stage.    

Second, to discover pathological inputs that trigger performance symptoms, {\em feedback metrics} must account for underlying systems-level metrics such as CPU time and memory usage. Prior work PerfFuzz~\cite{lemieux2018perffuzz} is limited to simply counting the number of exercised control flow graph edges as feedback (i.e., inducing a longer execution path). To trigger various performance skews in big data analytics, a user should be able to inject performance monitors that can watch partition-level runtime, peak memory usage, and garbage collection time and compare these metrics across multiple partitions. 

Third, in terms of {\em input mutation operations}, random bit-level input mutations used in traditional fuzzers like AFL~\cite{afl} may not be effective at exposing real performance problems. This is because the entire running time is strongly influenced by the {\em collective properties of the dataset} (e.g., key distribution, garbage collection that depends on the entire dataset's in-situ memory needs, differences in parallel task execution latency). In other words, input mutations for performance fuzzing must ensure that each record compiles with the input schema and that either the content of data set should be modified in ways that target specific performance symptoms. 

\noindent \textbf{Our Solution.} {\bf \tool} automatically generates test inputs to trigger a  performance skew. In {\tool}, a user can easily specify a performance symptom of interest using pre-defined performance monitors. {\tool} then automatically inserts the corresponding performance monitor and uses it as feedback guidance for grey-box fuzzing. To overcome the aforementioned challenges of adapting fuzz testing for performance workload generation\textemdash long latency, lack of performance feedback, and low-level input mutations, \tool combines three technical innovations:

First, to trigger a performance symptom appearing in the later computation stage, \tool uses a {\em phased fuzzing} approach. Deeper program paths cannot be easily reached through input generation; however, it is relatively easier to target a single user-defined function (UDF) in isolation. Our phased fuzzing targets an individual user-defined function at a given stage to gain knowledge about the intermediate inputs that trigger a performance symptom. Then using a pseudo-inverse function that is auto-generated by a large language model GPT-4~\cite{openai2023gpt4}, \tool converts the intermediate inputs to corresponding inputs at the beginning of the program, which are then used as improved seeds for fuzzing the entire program. 

Second, \tool enables users to specify performance symptoms by implementing customizable monitor templates which provide useful guidance metrics for fuzzing. \tool currently supports performance outlier detecting templates for computation, memory, and data skew symptoms. These templates relate symptoms with relevant performance metrics such as partition runtime, memory usage, and shuffle sizes.

Third, \tool improves its chances of constructing meaningful inputs with sets of skew-inspired mutations. \tool supports a variety of input mutation operations including data replication, column/field-level mutations for multi-value inputs, and mutations that exploit key-value distributions. 

\noindent {\bf Evaluation.} We compare the test generation time and the number of iterations required for triggering performance symptoms with and without phased fuzzing. Across our four case studies, \tool achieves more than \EvalResultSpeedupX speedup in time and requires less than \EvalResultIterPct iterations compared to \tool without phased fuzzing.
Additionally, \tool's template-inspired mutation probabilities result in a \EvalResultMutationSpeedup speedup in input generation time compared to a uniform sampling configuration.

Most fuzzing techniques focus on correctness testing with crash symptoms. Compared to crashes, performance problems are not explored much in automated fuzzing. \tool is the first automated fuzzer that can trigger data, memory, and computational skews.
\begin{itemize} 
\item We present an automated test generation framework that supports various kinds of performance monitors for big data analytics. \tool is built on Apache Spark and its key idea generalizes to other data-intensive scalable computing applications such MapReduce and Hadoop. 

\item We are the first to employ a {\em targeted, phased fuzzing} approach to expedite fuzz testing by generating intermediate inputs at a later stage first and then mapping them to the improved seeds by a pseudo-inverse function at the beginning of a program. It only takes up to 2 prompting iterations to generate the pseudo-inverse function and achieves \EvalResultSpeedupX compared to the baseline. 

\item We are the first to define skew-inducing input mutations for data-intensive analytics. These input mutations duplicate data rows, adjust key-value distributions, or add additional records with the same key while mixing values from other rows. These input mutations differ from typical bit-level mutations and are designed to trigger data, memory, and computation skews easily. 

\end{itemize}

%% file: motivating_collatz.tex
\newcommand{\solvedRDD}{\textit{solved}\xspace}

\section{Motivating Example} 
\label{sec:motivation}
\label{perfgen_motivating}

\begin{figure}[!th]
\vspace{-3ex}
\centering 
\input{code/code_collatz_simple}
\vspace{-3ex}
\caption{The \CollatzProgramName program which applies the \codefont{solve\_collatz} function to each input integer and sums the result by distinct integer input.}
\label{fig:code_collatz}
\vspace{-3ex}
\end{figure}

\begin{figure}[!th]
\centering 
\vspace{-2ex}
\input{code/code_collatz_solve_equation_udf}
\vspace{-3ex}
\caption{The \codefont{solve\_collatz} function used in Figure \ref{fig:code_collatz} to determine each integer's Collatz sequence length and compute a polynomial-time result based on the sequence length. For example, an input of \textit{3} has a Collatz length of 7 and calling \codefont{solve\_collatz(3)} takes 1 ms to compute, while an input of \textit{27} has a length of 111 and takes 4.9~s to compute.}
\label{fig:solve_collatz}
\vspace{-2ex}
\end{figure}

To demonstrate the challenges of performance debugging and how \perfgen addresses such challenges, we present a motivating example using a program inspired by \cite{collatz}. In this example, a developer uses the \CollatzProgramName program shown in Figure \ref{fig:code_collatz}. The \CollatzProgramName program consumes a string dataset of space-separated integers to compute a mathematical result for each distinct integer based on its \textit{Collatz} sequence length and number of occurrences. For each parsed integer, the program applies a mathematical function \codefont{solve\_collatz} (Figure \ref{fig:solve_collatz}) to compute a numerical result based on each integer's Collatz sequence length, in polynomial time with respect to that length. After applying \codefont{solve\_collatz} to each integer, the program then aggregates across each integer and returns the summed result per distinct integer.

Suppose the developer is interested in exploring the performance of this program, particularly the \solvedRDD variable which applies the \codefont{solve\_collatz} funtion. They want to generate an input dataset that will induce performance skew
by causing a single data partition to require at least five times the computation time of other partitions. In other words, they wish to find an input that meets the following symptom predicate:
\begin{center}
$\frac{\displaystyle Slowest Partition Runtime}{\displaystyle Second Slowest Partition Runtime} \ge 5.0$
\end{center}

As a starting point, the developer generates an initial input consisting of four single-record partitions: ``1", ``2", ``3", and ``4". However, this simple input does not result in any significant performance skew within the \CollatzProgramName program.

The developer initially turns to traditional fuzzing techniques for help in generating a skew-inducing input dataset. However, such approaches either flip individual bits or some bytes in the dataset in an attempt to produce new inputs. Because \CollatzProgramName's string inputs are parsed into integers, such inputs may not be program-compatible and therefore could not reach the \codefont{solve\_collatz} function and induce performance skew. Furthermore, traditional fuzzing techniques typically use code branch coverage as guidance for driving execution towards rare execution paths, and thus do not consider performance metrics. 

The developer decides to use \perfgen to generate an input that produces performance skew for the \CollatzProgramName program. Suppose that the developer suspects that \solvedRDD UDF may have a performance problem. \perfgen's phased fuzzing approach first targets this function to generate intermediate inputs. It then uses a pseudo-inverse function to convert the \solvedRDD's intermediate inputs to improved seeds for \CollatzProgramName. For example, Figure~\ref{fig:collatz_inverse_fn} shows such a pseudo-inverse function that guesses seed inputs from the intermediate inputs. Such a pseudo-inversion function does not need to be an exact inverse function as its goal is to generate improved seeds for subsequent fuzzing iterations. \perfgen applies the large language model to generate a pseudo-inverse function. As shown in Figure~\ref{fig:gpt-overview}, in our case, it took only 1 prompting to create this pseudo inverse function using ChatGPT.

\begin{figure}[t]
\centering 
\input{code/code_collatz_inverse_fn}
\vspace{-3ex}
\caption{
A pseudo-inverse function to convert \solvedRDD inputs into inputs for the entire \CollatzProgramName program (Figure \ref{fig:code_collatz}, lines 1-7).
}
\label{fig:collatz_inverse_fn}
\vspace{-3ex}
\end{figure}

Next, the developer defines their symptom in \perfgen by selecting a \textit{monitor template} and performance \textit{metric} from Tables \ref{table:metric_templates} and \ref{table:perf_metrics}. 
Based on the symptom predicate described earlier, they choose a \textit{NextComparison(5.0)} monitor template and the \textit{Runtime} metric to create the following symptom monitor. $[Runtime]$ is the collection of partition runtimes for a given job execution:
\begin{center}
    $\frac{\displaystyle max([Runtime])}{\displaystyle max([Runtime] - \{max([Runtime])\})} \ge 5.0$
\end{center}

\noindent This predicate inspects the partition runtimes for a given job execution and checks if the longest partition runtime is at least five times as long as all other partition runtimes.

Using this symptom monitor, \perfgen produces mutations from Table \ref{table:mutations} for both intermediate \solvedRDD inputs as well as \CollatzProgramName program inputs. For example, the mutations for \solvedRDD's \textit{(Int, Iterable[Int])} inputs include mutations which randomly replace the integer values in record keys or values, or alter the distribution of data by appending newly generated records. In addition to producing mutations, \perfgen also defines mutation sampling probabilities by assigning sampling weights to each mutation based on their alignment with the symptom definition; for example, mutations associated with computation skew have higher sampling probabilities when \perfgen is given a computation skew symptom.

\begin{figure}[t]
\begin{lstlisting}[language=Scala]
val config = PerfGenConfig(
    programOutput = sum, // HybridRDD output of entire program
    targetUDF = solved, // HybridRDD output of target UDF
    monitorTemplate = nextComparison, // Monitor Template / Symptom definition
    inputMutationMap, // Program mutations
    udfMutationMap, // UDF mutations
    seed = (("1"),("2"), ("3"), ("4")), // initial seed input
    inverse // pseudo-inverse function from Collatz program.
)
PerfGen.run(config)
\end{lstlisting}
    \vspace{-4ex}
    \caption{Code demonstrating how a user can configure \perfgen for the Collatz program}
    \label{code:perfgen_driver_example_collatz}
    \vspace{-2ex}
\end{figure}

\begin{figure*}[t]
    \centering
    \scalebox{1}{\includegraphics[width=\textwidth]{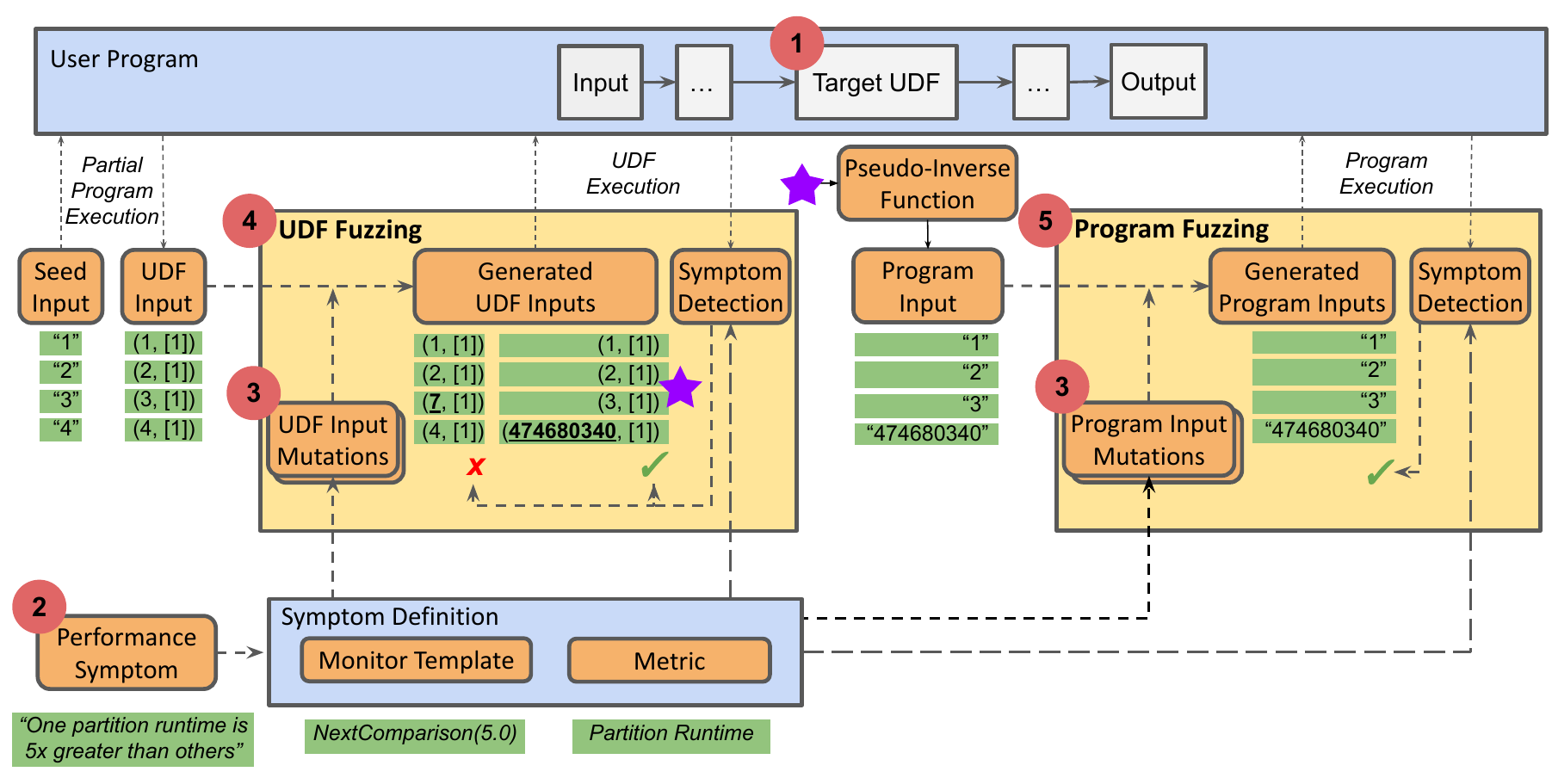}}
    \vspace{-4ex}
    \caption{An overview of \perfgen's phased fuzzing approach. Suppose that a user wants to generate test data that can induce a performance symptom of interest at a given stage.  
    \perfgen generates a set of mutations for both UDF and program input fuzzing. It first fuzzes the target UDF to reproduce the desired performance symptom. It then applies a pseudo-inverse function to generate improved seeds. These seeds are used to fuzz the entire program to induce the performance symptom of interest.}
    \label{fig:overview}
    \vspace{-4ex}
\end{figure*}

The user-specified target UDF, monitor template, and metric are shown in Figure \ref{code:perfgen_driver_example_collatz}. Using this configuration, \perfgen begins its phased fuzzing approach. It first executes \CollatzProgramName with the input data until reaching inputs to \solvedRDD, producing the partitioned UDF input shown in Figure \ref{fig:overview}. Next, it uses the derived mutations to fuzz \solvedRDD first. After a few iterations, it produces the symptom-triggering UDF inputs starred in Figure \ref{fig:overview} by adding the bolded record. As a result of the key's long Collatz length and the \codefont{solve\_collatz} function, this intermediate input executes slowly for only one of the data partitions and satisfies the performance skew definition.\footnote{
"474680340" has a Collatz sequence length of 192, while the remaining records' lengths are no more than 7.}

Next, \perfgen applies the pseudo-inverse function to this UDF input to produce the \CollatzProgramName program input shown in Figure \ref{fig:overview}. Upon testing, \perfgen finds that the converted inputs also exhibit performance skew for the full \CollatzProgramName and returns the dataset to the user for further analysis. At this point, the user now possesses a \CollatzProgramName program input triggering a performance skew symptom of interest.

%% file: code/code_collatz_simple.tex
\begin{lstlisting}[language=Scala]
val inputs = sc.textFile("collatz.txt") // read input
val  trips = inputs
  .flatMap(line => line.split(" ")) // split by space
  .map(s=>(Integer.parseInt(s),1))// parse ints into pair 
val groups=trips.groupByKey(4)//group into 4 partitions
val solved = groups.map { s => // apply UDF to generate
    (s._1, solve_collatz(s._1)) } //  new pair values
val sum=solved.reduceByKey((a, b) => a + b)//sum by key
\end{lstlisting}

%% file: code/code_collatz_solve_equation_udf.tex
\begin{lstlisting}[language=Scala]
def solve_collatz(m:Int): Int ={
    var k=m, i=0
    while (k>1) { // compute collatz sequence length, i
      i=i+1
      if (k % 2==0)  k = k/2 else {k=k*3+1}}
    var a=i+0.1
    for (j<-1 to i*i*i*i){ // O(i^4) computation loop
      a = (a + log10(a))*log10(a)}
    a.toInt
}
\end{lstlisting}

%% file: code/code_collatz_inverse_fn.tex
\begin{lstlisting}[language=Scala]
def inverse(udfInput: RDD[(Int, Iterable[Int])]): RDD[String] = {
    udfInput.flatMapValues(identity).map(s => s._1.toString)
}
\end{lstlisting}

%% file: approach.tex
\begin{figure*}[t]
\centering
   \scalebox{0.85}{\begin{minipage}{0.52\textwidth}
    \begin{subfigure}[t]{1\textwidth}
        \input{code/code_collatz_sparkrdd}
        \vspace{-1.5ex}
        \caption{A DISC Application \codefont{Collatz.scala}}
            \label{fig:collatz_sparkrdd}
    \end{subfigure}
    \end{minipage}}  
    \scalebox{0.85}{\begin{minipage}{0.55\textwidth}
    \begin{subfigure}[t]{1\textwidth}
        \vspace{4ex}
        \input{code/code_collatz_hybridrdd}
        \vspace{-3ex}
        \caption{Transformed \codefont{Collatz} with \codefont{HybridRDD}}
            \label{fig:collatz_hybridrdd}
            
        \begin{tikzpicture}[overlay]
          \draw[orange,thick,rounded corners] (7.2,1.4) rectangle (10.8,1.9) node[pos=.5] {\textcolor{black}{\bf\footnotesize User-defined function}};
        \end{tikzpicture}
    \end{subfigure}
    \end{minipage}} 
\vspace{-2ex}
\caption{\perfgen's \textit{HybridRDD} supports extraction and reuse of individual UDFs}
\vspace{-2ex}
\end{figure*}

\vspace{-1ex}
\section{Approach}
\label{perfgen_approach}

As shown in Figure~\ref{fig:overview}, \perfgen takes as input a DISC application built on Apache Spark, an initial input seed, and a symptom of interest defined using a monitor template. 

\perfgen extends a traditional fuzzing workflow with four novel contributions. Section~\ref{sec:udfisolation} describes   
\perfgen's \textit{HybridRDD} extension to the Spark RDD API in order to support execution of individual UDFs for more precise fuzzing.
Section~\ref{sec:perffeedback} enables a user to specify a desired symptom via execution \textit{metrics} and predefined \textit{monitor templates} which define patterns to detect symptoms. Section~\ref{sec:mutation} leverages type knowledge from the isolated UDF as well as the symptom definition to define a set of {\it skew-inspired mutations} designed to generate syntactically valid inputs geared towards producing the  performance symptom of interest. Finally, Section~\ref{sec:phasedfuzzing} combines these techniques to find symptom-reproducing UDF inputs first and to produce improved seeds for fuzzing the entire program end to end.

\begin{table*}[!th]
{    
    \linespread{0.5}\selectfont
    \centering
      \scalebox{0.85}{\input{tables/metric_templates}}
        
     \vspace{1ex}
    \caption{\textit{Monitor Templates} define \textit{predicates} that are used to (1) detect specific symptoms and (2) calculate feedback scores, given a collection of values $X$ derived using performance metrics definitions such as those from Table \ref{table:perf_metrics}. 
    }
        \label{table:metric_templates}
    \vspace{-2ex}
} 
\end{table*}

\subsection{Enabling an entry point to UDFs}
\label{sec:udfisolation}
DISC applications have longer latency than other applications, making them unsuitable for iterative fuzz testing end to end~\cite{zhang2020:bigfuzz}. Triggering a performance skew is also an extremely rare event to induce by chance via low-level mutation, when the symptom occurs deep in the program. Fuzzing is easier for a single UDF in isolation than fuzzing an entire application to reach a deep execution path. However, fuzzing a UDF requires having an entry point to the UDF. 

Suppose that a user desires to have a direct entry point to a specific UDF of interest (Figure \ref{fig:overview} label 1). Existing DISC systems such as Spark define datasets in terms of transformations (including UDFs) directly applied to previous datasets. As a result, such programs do not support decoupling UDFs from input datasets without manual refactoring or system modifications. It is nontrivial to execute a program (or subprogram) with new inputs.\footnote{For example, Spark's various \textit{RDD} implementations including \textit{MapPartitionsRDD} and \textit{ShuffledRDD} capture information about transformations via private, operator-specific objects such as iterator-to-iterator functions or Spark \textit{Aggregator} instances. Reusing these transformation definitions with new inputs requires direct access to Spark's internal classes.} 

To enable a direct entry point to a UDF, \perfgen wraps Spark \textit{RDD}s with its own \textit{HybridRDD}s. While \textit{HybridRDDs} are functionally equivalent to RDDs, they internally separate transformations from the datasets on which they are applied and store information about the corresponding input and output data types. 
Using this new \textit{HybridRDD} interface, a user can easily specify a target UDF of interest. Figures \ref{fig:collatz_sparkrdd} and \ref{fig:collatz_hybridrdd} illustrate the API changes required to leverage \perfgen's \textit{HybridRDD} for the \CollatzProgramName program discussed in Section \ref{sec:motivation}. \perfgen automatically decouples the \textit{map} transformation of \codefont{solved} from its predecessor (\codefont{grouped}) to produce a function of type \codefont{RDD[(Int, Iterable[Int])]} $=>$ \codefont{RDD[(Int, Int)]} which captures the \codefont{solve\_collatz} function used in the \codefont{map} transformation. 

\subsection{Modeling performance symptoms}
\label{sec:perffeedback}

\begin{table}[!th]
{   
    \centering
        \scalebox{0.8}{
\begin{tabular}{| l | l |}
\cellcolor{black}{\textcolor{white}{\at{\bf Performance Metric}}}
&\cellcolor{black}{\textcolor{white}{\at{\bf Skew\ Category}}}
\\\hline
Job Execution Time & Computation \& Data \\\hline
Garbage Collection Time & Memory \\\hline
Peak Memory Usage& Memory \\\hline
Memory Bytes Spilled on Disk & Memory \\\hline
Input Read Records Count& Data 
\\\hline
Output Write Records Count& Data \\\hline
Shuffle Read Records Count& Data \\\hline
Shuffle Read Bytes & Data \\\hline
Shuffle Write Records Count & Data \\\hline
Shuffle Write Bytes & Data 
\\\hline
\end{tabular}
        }        
    \caption{Performance metrics captured by \perfgen }
    \label{table:perf_metrics}
    \vspace{-3ex}
} 
\end{table}

Detecting performance skews often requires monitoring task execution time, the number of records read or written during a shuffle, and memory usage. To guide test generation towards exposing performance skews, \perfgen provides a set of eight customizable \textit{monitor templates} and ten performance \textit{metrics} that we constructed using Spark's Listener API. They are listed in Tables \ref{table:metric_templates} and \ref{table:perf_metrics} respectively. These metrics are measured at the level of each partition and each stage. 

Our insight behind these templates is that {\em DISC performance skews often follow patterns} and each performance symptom could be modeled using \textit{monitor template} with a choice of \textit{performance metrics} (Figure \ref{fig:overview} label 2). {\perfgen} uses these monitoring templates as both an oracle and a guidance feedback, as opposed to detecting crashes as an oracle and monitoring branch coverage for grey-box fuzzing. 

Consider a symptom where any partition's runtime during a program execution exceeds 100 seconds. This symptom can be defined by using the \textit{Runtime} metric and \textit{MaximumThreshold} monitor template, which evaluates the following predicate using the collected partition runtimes from Spark to determine if the performance symptom is triggered:
\vspace{-1ex}
\begin{center}
$\max([Runtime]) \ge 100 s$.
\end{center}
\vspace{-1ex}
\noindent In addition to detecting symptoms, the monitor template also provides a feedback score corresponding to the largest metric (runtime) value observed. 

While \perfgen models many symptoms via the definitions in Tables \ref{table:metric_templates} and \ref{table:perf_metrics}, other symptoms may require additional patterns or metrics. \perfgen enables users to define custom  templates by implementing a monitor template interface.
 
\subsection{Skew-Inspired Input Mutation Operations} \label{sec:mutation}
\begin{table*}[th]
{    
    \centering
        \scalebox{0.85}{\input{tables/mutations}}
    \vspace{1ex}
    \caption{ Skew-inspired mutation operations implemented by \perfgen for various data types and their typical skew categories. Some mutations depend on others (\eg due to nested data types).
    }
    \label{table:mutations}
} 
\vspace{-6ex}
\end{table*}
Consider the Collatz program from Section \ref{sec:motivation}, which parses strings as space-separated integers.  When bit-level or byte-level mutations are applied to such inputs, they can hardly generate meaningful data that drives the program to a deep execution path since bit-flipping is likely to destroy the data format or data type. For example, modifying an input "10" to "1a"
would produce a parsing error since an integer number is expected. Additionally, DISC applications include distributed performance bottlenecks such as data shuffling that is dependent on the characteristics of the entire dataset and may be difficult or impossible to trigger with only record-level mutations. Designing mutations to detect performance skews in DISC applications requires that (1) mutations must ensure type-correctness, and (2) mutations should be able to manipulate input datasets in ways that comprehensively exercise the performance-sensitive aspects of distributed applications. For example, mutations should shuffle or redistribute data.

For example, \perfgen targets data skew symptoms by defining mutations that alter the distribution of keys and values in tuple inputs, as well as mutations that extend the length of collection-based fields (which might be flattened into multiple records and contribute to data skew later in the application). \perfgen's mutations alter specific values or elements in tuple and collection datasets. \textit{AppendSameKey} mutation (M13 in Table \ref{table:mutations}) targets data skew by appending new records for a pre-existing key.

\perfgen adjusts the sampling probability of each mutation based on the skew category associated with the symptom of interest (Figure \ref{fig:overview} label 3). Table \ref{table:mutations} describes \perfgen's mutations that target different skew categories.

\vspace{-1ex}
\subsection{Phased Fuzzing} \label{sec:phasedfuzzing}
\begin{figure}[t]
\input{code/code_phased_fuzzing_outline_and_loop_merged}
\vspace{-3ex}
\caption{\perfgen's \textit{phased fuzzing} approach for generating symptom-reproducing inputs, using feedback scores from monitor templates to guide fuzzing.}
\vspace{-3ex}
\label{code:phased_fuzzing_and_loop}
\end{figure}
\perfgen's \textit{phased fuzzing} technique generates intermediate inputs to a given UDF and produces improved seeds for fuzzing the entire program, sketched in Figure \ref{code:phased_fuzzing_and_loop}.

\noindent\textbf{Step 1. UDF Fuzzing.}
\perfgen generates an initial UDF input by partially executing the original program. Using this intermediate result as a seed, it then fuzzes the target UDF using the procedure outlined in Figure \ref{code:phased_fuzzing_and_loop}. The process is illustrated in Figure \ref{fig:overview} label 4 with concrete inputs from the motivating example.
Two nontrivial outcomes exist for each fuzzing loop iteration: (1) the monitor template detects that the desired symptom is triggered and terminates the fuzzing loop or (2) the monitor template does not detect skew but returns a feedback score that is better than previously observed, so \perfgen adds saves the mutated input, updates the best observed feedback score, and resumes fuzzing with the updated input queue.

\noindent\textbf{Step 2. Pseudo-Inverse Function generation with the Large Language Model.}
While targeted UDF fuzzing enables \perfgen to generate symptom-triggering intermediate inputs, the final objective is to identify inputs to the entire program. \perfgen uses a \textit{pseudo-inverse function} to convert intermediate UDF inputs to improved seeds. Such pseudo-inverse function does not need to be an exact inverse function as its goal is to bootstrap seed input generation. \CollatzProgramName pseudo-inverse function is shown in Figure \ref{fig:collatz_inverse_fn}.

\begin{figure}[t]
    \centering
    \scalebox{0.5}{\includegraphics[width=\textwidth]{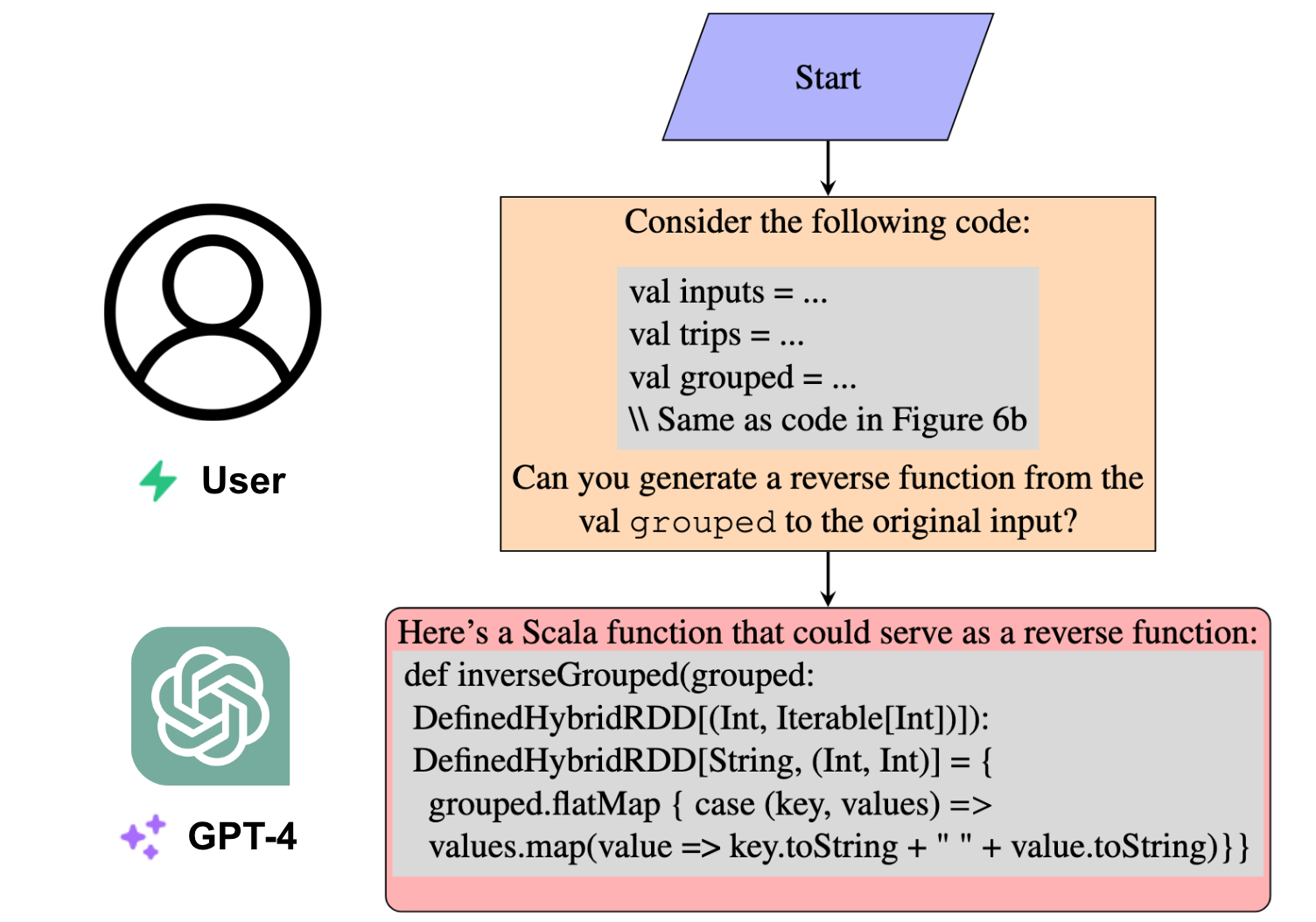}}
    \vspace{-2ex}
    \caption{An example of \perfgen's pseudo-inverse function generation.}
    \label{fig:gpt-overview}
    \vspace{-5ex}
\end{figure}

As illustrated in Figure~\ref{fig:gpt-overview}, a pseudo-inverse function can be generated by a large language model such as GPT-4. For example, \perfgen prompts the GPT-4 model to generate an inverse function  by submitting the source code of \CollatzProgramName; the request could be framed as, ``Could you generate an inverse function that transitions from the RDD \codefont{grouped} to the original input? There is no need to provide the precise reverse function, just try your best.'' In response, the AI model will strive to deliver an appropriate inverse function.

Take, for instance, a dataset comprising student grades with an aggregation that computes the mean grade per course. Given an intermediate dataset displaying courses and their average grades, GPT-4  ynthesizes a pseudo-inverse function that outputs a single student grade per course, rounded to the nearest integer equivalent of the average grade. While such a function can only approximate average grades and consequently may not precisely replicate the provided intermediate inputs, the function can still facilitate the production of enhanced seeds.

A pseudo-inverse functions for the portion of a program that precedes the target User Defined Function (UDF), is independent of the target UDF implementation. For example, Figure \ref{fig:collatz_inverse_fn} for the \CollatzProgramName program in Figure \ref{fig:code_collatz} excludes the target UDF \codefont{solve\_collatz}. Additionally, a pseudo-inverse function remains unaffected by target symptoms. Therefore, it is external to \perfgen's \textit{phased fuzzing} process and in most cases, is simple enough for a large language model to automatically generate an accurate version. In our evaluation, pseudo-inverse functions averaged 6 lines of code.

\noindent\textbf{Step 3. End-to-End Fuzzing with Improved Seeds.} \label{sec:program_fuzzing}
As a final step, \perfgen tests the pseudo-inverse function result to see if it is a symptom-triggering input. If not, it uses the derived program input as an improved seed for fuzzing the entire application as shown in Figure \ref{fig:overview} label 5. This step resembles UDF fuzzing (Figure \ref{code:phased_fuzzing_and_loop}) and reuses the same monitor template, but initializes with the pseudo-inverse function output as a seed and utilizes a different set of mutations suitable for the entire program's input data type. \perfgen achieves at least 11x speedup compared to a traditional fuzzing approach when generating inputs to trigger performance symptoms in our case studies.

%% file: code/code_collatz_sparkrdd.tex
\lstset{language=C++}
\begin{lstlisting}[style=MyJavaSmallStyle]
val !\colorbox{white!70!cyan}{inputs = sc.textFile}!("collatz.txt")
val  !\colorbox{white!80!red}{trips =}! inputs
  .flatMap(line => line.split(" "))
  .map(s=>(Integer.parseInt(s),1))
val !\colorbox{white!80!green}{grouped =}! trips.groupByKey(4)
val !\colorbox{white!70!orange}{solved =}! grouped.map { s => 
    (s._1, solve_collatz(s._1)) }
val !\colorbox{white!85!black}{sum =}! solved.reduceByKey((a, b) => a + b)
\end{lstlisting}

%% file: code/code_collatz_hybridrdd.tex
\lstset{language=C++}
\begin{lstlisting}[style=MyJavaSmallStyle]
val !\colorbox{white!70!cyan}{inputs = HybridRDD(sc.textFile}!("collatz.txt"))
val  !\colorbox{white!80!red}{trips: HybridRDD[String, (Int, Int)] =}! inputs
 .flatMap(line => line.split(" "))
 .map(s=>(Integer.parseInt(s),1))
val !\colorbox{white!80!green}{grouped: HybridRDD[(Int, Int), (Int, Iterable[Int])] =}!
 trips.groupByKey(4)
// RDD corresponding to the target UDF
!\colorbox{white!70!orange}{\textcolor{blue}{val} solved: HybridRDD[(Int, Iterable[Int]), (Int, Int)] = }!
!\colorbox{white!70!orange}{ grouped.map \{ s =>(s.\_1, solve\_collatz(s.\_1)) \}}!
val !\colorbox{white!85!black}{sum: HybridRDD[(Int, Int), (Int, Int)] =}! solved.reduceByKey((a, b) => a + b)
\end{lstlisting}

%% file: tables/metric_templates.tex
 
\begin{threeparttable}
\begin{tabular}{| l | c | p{0.68\linewidth}|} 
\cellcolor{black}{\textcolor{white}{\at{\bf Template(parameters)}}}
&\cellcolor{black}{\textcolor{white}{\at{\bf Predicate}}}
&\cellcolor{black}{\textcolor{white}{\at{\bf Description}}}
\\\hline
MaximumThreshold($X$,$t$) & \makecell{$\max(X) \ge t$, where $t$ = value threshold} & Compares the maximum value of $X$ to a threshold $t$.
\\\hline
NextComparison($X$,$t$) & \makecell{$\dfrac{\max(X)} {\max(X - \{\max(X)\})} \ge t$,\\where $t$ = ratio threshold} & Computes the ratio between the two largest metric values in $X$  and compares it to a threshold $t$.
\\\hline
IQROutlier($X$,$t$) & \makecell{$\dfrac{\max(\max(X) - Q_3, Q_1 - \min(X))}{Q_3 - Q_1} \ge t$,\\where $Q_1,Q_3$ = first and third quartiles of $X$,\\$t$ = IQR distance threshold (default 1.5)} & Computes the largest interquartile range (IQR) \cite{IQR} distance in $X$ and compares it to a threshold $t$.
\\\hline
Skewness($X$,$t$) & \makecell{$\dfrac{m_3}{\sigma^3} \ge t$,\\where $m_3$ = third central moment of $X$,\\$\sigma$ = standard deviation of $X$,\\$t$ = skewness threshold (default 1.0)} & Computes the skewness \cite{Skewness} of $X$ and compares it to a threshold $t$.
\\\hline
ZScore($X$,$t$) & \makecell{$\dfrac{\max(X) - \mu}{\sigma} \ge t$,\\where $\mu$ = mean of $X$,\\$\sigma$ = standard deviation of $X$,\\ $t$ = z-score threshold} & Computes the largest z-score \cite{ZScore} in $X$ and compares it to a threshold $t$.
\\\hline
ModZScore($X$,$t$) & \makecell{$\dfrac{\max(X) - M}{1.486*MAD} \ge t$,\\where $M$ = median of $X$,\\$MAD$ = median absolute deviation of $X$,\\$t$ = modified z-score threshold} & Computes the largest modified z-score \cite{ModZScore} in $X$ and compares it to a threshold $t$.
\\\hline
LeaveOneOutRatio($X$,$t$) & \makecell{$\dfrac{\max(X)}{\text{mean}(X - \{\max(X)\})} \ge t$,\\where $t$ = target ratio threshold} & Computes the ratio between the largest metric and the average of all other metrics, and compares it to a threshold $t$.
\\\hline
ErrorDetection($X$,$s$, $mt$) & \makecell{error is thrown and \\error message contains substring $s$} & Monitors for thrown exceptions with error messages containing the specified substring $s$. An underlying monitor template $mt$ is required to provide a feedback score during fuzzing.
\\\hline
\end{tabular}
\end{threeparttable}

%% file: tables/mutations.tex
\begin{tabular}{| c | c | c | c | p{0.55\linewidth} |}
\cellcolor{black}{\textcolor{white}{\at{\bf ID}}}
& \cellcolor{black}{\textcolor{white}{\at{\bf Name}}}
& \cellcolor{black}{\textcolor{white}{\at{\bf Data\ Type}}}
&\cellcolor{black}{\textcolor{white}{\at{\bf Target\ Skew(s)}}}
&\cellcolor{black}{\textcolor{white}{\at{\bf Description}}}
\\\hline
\hspace{-1ex}
M1 & ReplaceInteger & Integer & Computation & Replace the input integer with a randomly generated integer value within a configurable range (default: [0, \codefont{Int.MaxValue})).
\\\hline
M2 & ReplaceDouble & Double & Computation & Replace the input double with a randomly generated double value within a configurable range (default: [0, \codefont{Double.MaxValue}))
\\\hline
M3 & ReplaceBoolean & Boolean & Computation & Replace the input boolean with a random boolean value.
\\\hline
M4 & ReplaceSubtring & String & Computation & Mutate a string by replacing a random substring (including either empty or the full string) with a newly generated random string of random length within a configurable range (default: [0, 25)).
\\\hline
M5 & ReplaceCollectionElement & Collection & Computation & Randomly select and mutate a random element within a collection according to its type.
\\\hline
M6 & AppendCollectionCopy & Collection & \makecell{Computation,\\ Data, Memory} & Extend a collection by appending a copy of itself.
\\\hline
M7 & ReplaceTupleElement & 2-Element Tuple & Computation & Randomly mutate an element within a two-element tuple according to its type.
\\\hline
M8 & ReplaceTripleElement & 3-Element Tuple & Computation & Randomly mutate an element within a three-element tuple according to its type. 
\\\hline
M9 & ReplaceQuadrupleElement & 4-Element Tuple & Computation & Randomly mutate an element within a four-element tuple according to its type.
\\\hline
M10 & ReplaceRandomRecord & Dataset & Computation & Randomly select a record and mutate it according to one of the mutations applicable to the dataset type. For example, this mutation could choose a random integer out of an integer dataset and apply the \textit{ReplaceInteger} mutation. 
\\\hline
M11 & PairKeyToAllValues & 2-Element Tuple Dataset & Data, Memory & Randomly select a random record. For each distinct value within that record's partition, append a new record to the partition consisting of the the selected record's key and the distinct value, such that the key is paired with every value in the partition.
\\\hline
M12 & PairValueToAllKeys & 2-Element Tuple Dataset & Data & Similar to \textit{PairKeyToAllValues} but instead pairing a random record's value with all distinct keys in a partition.
\\\hline
M13 & AppendSameKey & 2-Element Tuple Dataset & Data, Memory & Randomly select a random record. Append additional records consisting of that record's key paired with mutations of its value some number of times (default: up to 10\% of partition size).
\\\hline
M14 & AppendSameValue & 2-Element Tuple Dataset & Data & Similar to \textit{AppendSameKey} but instead with a fixed value and mutated keys.
\\\hline
\end{tabular}

%% file: code/code_phased_fuzzing_outline_and_loop_merged.tex
\begin{lstlisting}[language=Scala]
def phasedFuzzing[I, U, O](conf: PerfGenConfig[I, U, O]): RDD[I] = {
// Step 1: Fuzz the target UDF to produce symptom-triggering intermediate inputs
	// partially run program up until UDF
	val udfSeed: RDD[U] = computeUDFInput(conf.seed) 
	val udfSymptomInput: RDD[U] = fuzz(conf.udfProgram,
	  udfSeed, conf.monitorTemplate, conf.udfMutations)
// Step 2: Use pseudo-inverse function to generate program seed
	val programSeed: RDD[I] = conf.inverseFn.apply(udfSymptomInput)
// Step 3: Fuzz the full program to produce symptom-triggering program inputs
	val programSymptomInput: RDD[I] = fuzz(conf.fullProgram, programSeed, conf.monitorTemplate, conf.programMutations)
	return programSymptomInpu}

def fuzz[T,U](progFn: RDD[T] => RDD[U], seed: RDD[T], monitor: MonitorTemplate, mutations: MutationMap[T]) = {
	val seeds = List(seed)
	var maxScore = 0.0
	while(true) { // not timed out
		// select a seed and apply a mutation
		val base = sample(seeds)
		val newInput = mutations.sample().apply(base)
		val programOutput = progFn(newInput) // test the new input
		// Use monitor template to check if symptom was reproduced
		// or if feedback score was increased.
		val metrics = config.metric.getLastExecutionMetrics()
		val (meetsCriteria, feedbackScore) = monitor.checkSymptoms(metrics)
		if(meetsCriteria) { // symptom reproduced
			return newInput}
		else if (feedbackScore>maxScore) {// score increased
			maxScore = feedbackScore
			seed.append(newInput)}}}
\end{lstlisting}

%% file: evaluations_2022.tex
\newcommand{\DeptProgramName}{\textit{DeptGPAsMedian}\xspace}
\section{Evaluation}

\label{sec:evaluation}
\label{perfgen_evaluation}

We evaluate \perfgen with the following research questions:

\begin{description}

\item[RQ1] {How much speedup in total execution time can \perfgen achieve by phased fuzzing?} 

 \item[RQ2] {How much reduction in the number of fuzzing iterations does \perfgen provide through improved seeds derived from phased fuzzing?}

\item[RQ3] {How much improvement in speedup is gained by \perfgen's adjustment of mutation sampling probabilities based on the target symptom?} 
\end{description} 

RQ1 assesses overall time savings in using \perfgen, while RQ2 measures the reduction in the number of required fuzzing iterations. RQ3 explores the effects of mutation sampling probabilities on test input generation time.

\noindent{\bf Evaluation Setup}.
 As a baseline, we compare against a simplified version of \perfgen that does not apply phased fuzzing.
 This is because existing dataflow application fuzzers such as BigFuzz \cite{zhang2020:bigfuzz} are unable to detect performance symptoms or monitor underlying performance characteristics of Spark programs. The baseline configuration fuzzes the original program with the same monitor template but invokes the entire program with the initial seed input. Similar to the \perfgen setup, the baseline fuzzes the program until a skew-inducing input is identified. As pseudo-inverse functions are not tied to a specific symptom and can be potentially reused, we do not include their derivation times in our results; in practice, we found that each pseudo-inverse function definition required no more than five minutes to implement.

Each evaluation is run for up to four hours, using Spark 2.4.4's local execution mode on a single machine with 16GB RAM and 2.6 GHz 6-core Intel Core i7 processor.

\input{case_studies/case_study_collatz}

\input{case_studies/case_study_wordcount}

\begin{table*}[th]
{    
    \linespread{1.0}\selectfont
    \centering
        \scalebox{0.8}{\input{tables/case_study_results}}
        
    \vspace{1ex}
    \caption{Fuzzing times and iterations for each case study program. For programs marked with a "*", the baseline evaluation timed out after 4 hours and was unsuccessful in reproducing the desired symptom.}
    \label{table:case_study_results}
} 
       \vspace{-3ex}
\end{table*}

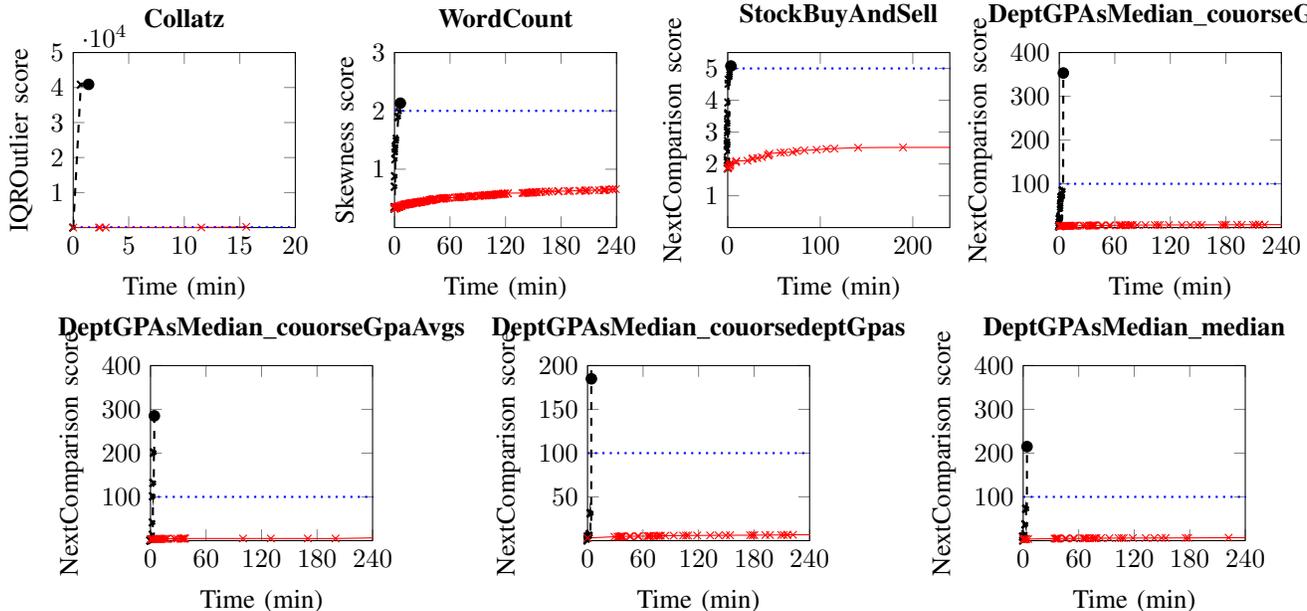
\begin{figure*}[!th]
    \centering
    \input{plots/eval_time_series_4x1}
    \vspace{-2ex}
    \caption{Time series plots of each case study's monitor template feedback score against time. \perfgen results are plotted in black with the final program result indicated by a circle, while baseline results are plotted in red crosses. The target threshold for each case study's symptom definition is represented by a horizontal blue dotted line.
    }
\label{fig:time_series} 
\vspace{-2ex}
\end{figure*}

\input{case_studies/case_study_stocks}

\input{case_studies/case_study_dept}

\subsection{Improvement in RQ1 and RQ2}

Table \ref{table:case_study_results} presents each case study's evaluation results, and Figure \ref{fig:time_series} shows each case study's progress over time. Averaged across all four case studies,\footnote{As three of the four case study baselines timed out after four hours, numbers are reported as bounds.} \perfgen leads to a speedup of at least \EvalResultSpeedupX while requiring no more than \EvalResultIterPct of the program fuzzing iterations required by the baseline. Additionally, \perfgen's UDF fuzzing process accounts for an average \EvalResultPhTimePct of its total execution time.

\subsection{RQ3: Effect of Mutation Weights}
\begin{figure}[t]
    \centering
    \input{plots/rq3_dept_gpa_time}
    \vspace{-4ex}
    \caption{Plot of \perfgen input generation time against varying sampling probabilities for the \textit{M13} and \textit{M14} mutations used in the \textit{DeptGPAsMedian} program.}
\label{plot:rq3_dept_gpa}
\vspace{-3ex}
\end{figure}
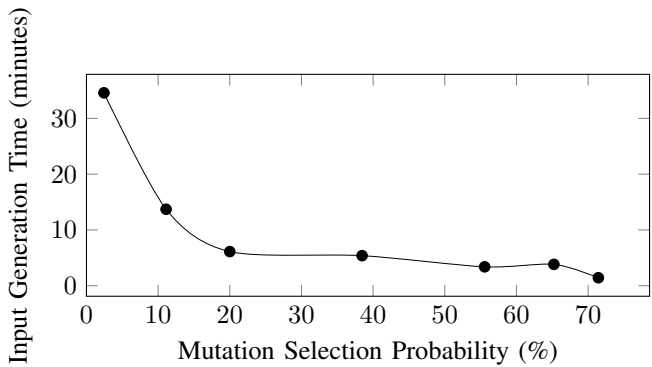

Using the \DeptProgramName program, we experiment with the mutation sampling probabilities to evaluate their impact on \perfgen's ability to generate symptom-triggering inputs. We reuse the same program, monitor template, and performance metric as in the case study (Section \ref{sec:case_study_deptgpas}), but vary the weight of the \textit{M13} and \textit{M14} mutations. As discussed in section \ref{sec:mutation}, mutation sampling probabilities are determined by weighted random sampling. In addition to the original weight of 5.0 in the case study, we also experiment with weights of 0.1, 0.5, 1.0, 2.5, 7.5, and 10.0 which result in individual mutation probabilities ranging from 2.44\% to 71.43\%. For each value, we average over 5 executions and report the total time required for \perfgen to generate an input that triggers the original \DeptProgramName symptom. 

Execution times for each sampling weight are plotted in Figure \ref{plot:rq3_dept_gpa}. We find that \perfgen's template-dependent weight of 5.0 leads to a speedup of \EvalResultMutationSpeedup compared to a configuration in which no extra weight is assigned (i.e., uniform weights of 1.0).
More generally, we also observe that the total time required to generate a satisfying input appears to be inversely proportional to the weights of the aforementioned mutations. 

In total, the range of execution times for each of the evaluated sampling weights ranged between 23.32\% and 564.74\% of the time taken for an unweighted evaluation.
Compared to an unweighted scenario, reducing the same weights to 0.5 and 0.1 resulted in 123.97\% and 464.74\% increases in total time.

%% file: case_studies/case_study_collatz.tex

\subsection{Case Study: Collatz Conjecture}
The \CollatzProgramName case study is based on the description in Section \ref{sec:motivation}. It parses a dataset of space-separated integers and applies a \textit{Collatz}-sequence-based mathematical function to each integer.  This case study's symptom definition differs from that in Section \ref{sec:motivation}, while other details including pseudo-inverse function and generated datasets remain the same.

The developer is interested in inputs that will exhibit severe computation skew in which one outlier partition takes more than 100 times longer to compute than others due to the \codefont{solve\_collatz} function. As this function is called in the transformation that produces \solvedRDD variable, they specify \solvedRDD as  function of interest for \perfgen's phased fuzzing. The developer defines their performance symptom by using the \textit{Runtime} metric with an \textit{IQROutlier} monitor template, specifying a target threshold of 100.0.

Based on Section~\ref{sec:mutation}, \perfgen produces the following mutations and weights for \solvedRDD variable and specified skew symptom:

\begin{center}
\centering 
\scalebox{0.80}{
\begin{tabular}{|l|c|r|}
\cellcolor{black}{\textcolor{white}{\at{\bf Mutation~Operators}}}
& \cellcolor{black}{\textcolor{white}{\at{\bf Weight}}}
& \cellcolor{black}{\textcolor{white}{\at{\bf Sampling~\%}}}
\\\hline
M10 + M7 + M1 & 1.0 & 11.1\%
\\\hline
M10 + M7 + M5 + M1 & 5.0 & 55.5\%
\\\hline
M10 + M7 + M6 & 1.0 & 11.1\%
\\\hline
M11 & 0.5 & 5.6\%
\\\hline
M12 & 0.5 & 5.6\%
\\\hline
M14 & 1.0 & 11.1\%
\\\hline
\end{tabular}
\par
}
\end{center}

\perfgen begins to generate test inputs and produces the datasets shown in Figure~\ref{fig:overview}. Its UDF fuzzing phase requires 3 iterations and 41~seconds. Its end-to-end fuzzing phase requires no iterations after the pseudo-inverse function is applied, as improved seeds immediately trigger the symptom of interest.

For comparison, we also evaluate the \CollatzProgramName program under our baseline configuration and it produces a symptom-triggering input after 12,166 iterations and 15.6~min by changing the "4" record to "338", which has a \textit{Collatz} length of 50.

\CollatzProgramName evaluation results are summarized in Table \ref{table:case_study_results}, with the progress of \textit{IQROutlier} feedback scores plotted in Figure \ref{fig:time_series}. Compared to the baseline, \perfgen's approach produces a 11.17X speedup and requires 0.008\% of the program fuzzing iterations. Additionally, \perfgen spends 49.14\% of its total input generation time on the UDF fuzzing process. While both configurations can successfully generate inputs that trigger the performance symptom of interest, \perfgen can do more efficiently because its type knowledge allows it to focus on generating integer inputs. The baseline is restricted to string-based mutations which often fail the integer parsing process.
\vspace{-2ex}

%% file: case_studies/case_study_wordcount.tex
\subsection{Case Study: WordCount}
\newcommand{\WordCountProgramName}{\textit{WordCount}\xspace}
\newcommand{\countsRDD}{\textit{counts}\xspace}

Suppose a developer is interested in the \WordCountProgramName program from \cite{BigDataBench}
.
\WordCountProgramName reads a dataset of Strings and counts how often each space-separated word appears in the dataset. As a starting input dataset, the developer uses a 5MB sample of Wikipedia entries consisting of 49,930 records across 20 partitions.


\begin{figure}[ht]
\centering 
\input{code/code_wordcount_program}
\vspace{-2ex}
\caption{The \WordCountProgramName program implementation in Scala which counts the occurrences of each space-separated word.}
\vspace{-2ex}
\label{fig:code_wordcount}
\end{figure}



The developer wants to generate an input for which the number of shuffle records written per partition exhibits a statistical skew value of more than 2. 
They identify the RDD corresponding to summation operation (implemented by \textit{reduceByKey(\_+\_)}) \countsRDD variable on line 4 in Figure \ref{fig:code_wordcount} 
as the UDF of interest because it induces a data shuffle, and
define the desired symptom by using the \textit{Shuffle Write Records} metric in combination with a \textit{Skewness} monitor template with a threshold of 2.0.

The target UDF
takes as input tuples of the type \textit{(String, Integer)}. Expecting a large number of intermediate records, the developer chooses to configure \perfgen to use a decreased duplication factor of 0.01. As the integer values are fixed to $1$, the developer also disables mutations which modify values in the UDF inputs. In addition to these configurations, \perfgen uses the data skew symptom to produce the following mutations and adjusts their sampling weights to bias towards producing data skew:


    


\begin{center}
\centering 
\scalebox{0.80}{\centering
\begin{tabular}{|l|c|r|}
\cellcolor{black}{\textcolor{white}{\at{\bf Mutation~Operators}}}
& \cellcolor{black}{\textcolor{white}{\at{\bf Weight}}}
& \cellcolor{black}{\textcolor{white}{\at{\bf Sampling~\%}}}
\\\hline
M10 + M7 + M4 & 1.0 & 14.3\%
\\\hline
M12 & 1.0 & 14.3\%
\\\hline
M14(duplicationFactor = 0.01) & 5.0 & 71.4\%
\\\hline
\end{tabular}
\par
}
\end{center}

Finally, PERFGEN generates a suitable pseudo-inverse function with GPT-4. As there is no way to reliably reconstruct the original strings from the tokenized words, the large language model implements a simple function that constructs input lines from consecutive groups of up to 50 words.
\begin{minipage}{\linewidth}
\input{code/code_wordcount_inverse_fn}
\end{minipage}
 
\perfgen executes \WordCountProgramName with the provided input dataset up until the target UDF generates a UDF input seed consisting of each word paired with a "1". It then applies mutations to this input until it generates a symptom-triggering UDF input after 37s and 357 iterations. Next, \perfgen applies the pseudo-inverse function to the input from UDF Fuzzing to produce an input for the full \WordCountProgramName program. It then tests this input and finds that the symptom is triggered, so no additional program fuzzing iterations are required.

For comparison, we evaluate \WordCountProgramName under the baseline configurations specified earlier in Section \ref{sec:evaluation}, using the same sample of Wikipedia data. The baseline times out after approximately 4 hours and 46,884 iterations without producing any inputs that trigger the target symptom.

Table \ref{table:case_study_results} summarizes the \WordCountProgramName evaluation results, and Figure \ref{fig:time_series} visualizes the progress of the maximum attained skewness statistics determined by the \textit{Skewness} monitor template. 
Compared to the baseline which is unable to produce results after 4 hours, \perfgen produces a speedup of at least 37.43X while requiring at most 0.0002\% of the program fuzzing iterations. 
98.48\% of \perfgen's total execution time is spent on UDF fuzzing.

While \perfgen is able to meet the target skewness threshold of 2.0, the baseline times out while never exceeding a skewnesss of 0.7. This gap in skewness comes from the baseline's inability to produce large quantities of new words which directly contribute to the number of shuffle records written by Spark. Meanwhile, \perfgen's \textit{M14} mutation produces many distinct words in each iteration, and thus enables \perfgen to quickly trigger the target symptom.
\vspace{-2ex}

%% file: code/code_wordcount_program.tex
\begin{lstlisting}[language=Scala]
val inputs = HybridRDD(sc.textFile("wiki_data")))
val words = inputs.flatMap(line => line.split(" "))
val wordPairs = words.map(word => (word, 1))
val counts = wordPairs.reduceByKey(_ + _)
\end{lstlisting}

%% file: code/code_wordcount_inverse_fn.tex
\begin{lstlisting}[language=Scala]
def inverse(udfInput: RDD[(String, Int)]): RDD[String]
 = {val words = udfInput.map(_._1)
    words.mapPartitions(wordIter => 
        wordIter.grouped(50).map(_.mkString(" ")))}
\end{lstlisting}

%% file: tables/case_study_results.tex
\begin{tabular}{| c | R{3em} R{3.5em} R{3em} | R{3.5em} R{3.5em} R{2.6em} | R{4em} | R{4em} R{3em} | R{3.2em} R{3.2em} R{3.2em} |}
\hline

\multicolumn{1}{|c|}{\multirow{7}{*}{\bf Program}}
& \multicolumn{3}{c|}{\multirow{2}{*}{\bf UDF Fuzzing}}
& \multicolumn{3}{c|}{\multirow{2}{*}{\bf Program Fuzzing}}
& \multicolumn{1}{c|}{\multirow{2}{*}{\parbox{3.8em}{\centering \bf \perfgen\ Total}}}
& \multicolumn{2}{c|}{\multirow{2}{*}{\bf Baseline}}
& \multicolumn{3}{c|}{\multirow{2}{*}{\bf \perfgen\ vs.\ Baseline}} \\
& & & & & & & & & & & & \\
\cline{2-13}

& \multirow{4}{*}{\parbox{3em}{\centering \bf Seed Init.\\\bf (ms)}}
& \multirow{4}{*}{\parbox{3.5em}{\centering \bf Duration\\\bf (ms)}}
& \multirow{4}{*}{\parbox{3.2em}{\centering \bf \#\ Iter.}}
& \multirow{4}{*}{\parbox{3.5em}{\centering \bf P-Inv.\\\bf Func.\\\bf Appl.\ (ms)}}
& \multirow{4}{*}{\parbox{3.5em}{\centering \bf Duration\\\bf (ms)}}
& \multirow{4}{*}{\parbox{2.6em}{\centering \bf \#\ Iter.}}
& \multirow{4}{*}{\parbox{4em}{\centering \bf Duration\\\bf (ms)}}
& \multirow{4}{*}{\parbox{4em}{\centering \bf Duration\\\bf (ms)}}
& \multirow{4}{*}{\parbox{3em}{\centering \bf \#\ Iter.}}
& \multirow{4}{*}{\parbox{3.2em}{\centering \bf Speedup}}
& \multirow{4}{*}{\parbox{3.2em}{\centering \bf Iter. \%\\\bf Program\\Fuzzing}}
& \multirow{4}{*}{\parbox{3.3em}{\centering \bf Time \%\\\bf Phased\\Fuzzing}}
\\

& & & & & & & & & & & & \\
& & & & & & & & & & & & \\
& & & & & & & & & & & & \\
\hline

Collatz & 1,259 & 41,221 & 3 & 310 & 41,095 & 1 & 83,888 & 937,071 & 12,166 & 11.17 & 0.008\% & 49.14\% \\

WordCount* & 4,299 & 378,946 & 357 & 986 & 544 & 1 & 384,778 & 14,401,990 & 46,884 & 37.43 & 0.002\% & 98.48\% \\

StockBuyAndSell* & 1,450 & 205,084 & 4,775 & 601 & 208 & 1 & 207,346 & 14,402,428 & 40,010 & 69.46 & 0.002\% & 98.91\% \\

DeptGPAsMedian\_courseGpas* & 2,282 & 259,205 & 1,519 & 736 & 638 & 1 & 262,864 & 14,405,503 & 21,575 & 54.80 & 0.005\% & 98.61\% \\

DeptGPAsMedian\_courseGpaAvgs* & 2,616 & 251,612 & 1,306 & 636 & 638 & 1 & 255,502 & 14,405,503 & 21,575 & 56.38 & 0.005\% & 98.23\% \\

DeptGPAsMedian\_deptGpas* & 1,707 & 271,538 & 2,306 & 107 & 638 & 1 & 273,990 &  14,405,503 &  21,575 & 52.57 &0.005\% & 98.09\% \\

DeptGPAsMedian\_median* & 881 & 292,555 & 1,387 & 852 & 638 & 1 & 294,926 & 14,405,503 & 21,575 & 49.02& 0.005\% & 97.96\% \\
\hline

\end{tabular}

%% file: plots/eval_time_series_4x1.tex
    \begin{tikzpicture}
    \begin{axis}[title=\textbf{Collatz},xlabel={Time (min)},ylabel={IQROutlier score},ylabel near ticks,
    xmin=0,xmax=20,xmode=linear,
    ymax=50000,ymin=0,ytick={10000,20000,...,50000},
    width=0.25\textwidth]
      \input{plots/time-series-subplots/collatz_timeseries}
    \end{axis}
    
    \end{tikzpicture}
    \begin{tikzpicture}
    \begin{axis}[title=\textbf{WordCount},xlabel={Time (min)},ylabel={Skewness score},ylabel near ticks,
    xmin=0,xmax=240,xtick={0,60,120,...,240},xmode=linear,ymax=3,ymin=0,ytick={1,2,3},width=0.25\textwidth]
       \input{plots/time-series-subplots/wordcount_timeseries}
    \end{axis}
    \end{tikzpicture}
    \begin{tikzpicture}
    \begin{axis}[title=\textbf{StockBuyAndSell},xlabel={Time (min)},ylabel={NextComparison score},ylabel near ticks,
    xmin=0,xmax=240,xmode=linear,ymax=5.5,ymin=0,ytick={1,2,...,5},width=0.25\textwidth]
       \input{plots/time-series-subplots/stockbuysell_timeseries}
    \end{axis}
    \end{tikzpicture}
    \begin{tikzpicture}
    \begin{axis}[title=\textbf{DeptGPAsMedian\_couorseGpas},xlabel={Time (min)},ylabel={NextComparison score},ylabel near ticks,
    xmin=0,xmax=240,xtick={0,60,120,...,240},xmode=linear,ymax=400,ymin=0,ytick={100,200,...,400},width=0.25\textwidth]
       \input{plots/time-series-subplots/deptgpasmedian_timeseries}
    \end{axis}
    \end{tikzpicture}
    \begin{tikzpicture}
    \begin{axis}[title=\textbf{DeptGPAsMedian\_couorseGpaAvgs},xlabel={Time (min)},ylabel={NextComparison score},ylabel near ticks,
    xmin=0,xmax=240,xtick={0,60,120,...,240},xmode=linear,ymax=400,ymin=0,ytick={100,200,...,400},width=0.25\textwidth]
       \input{plots/time-series-subplots/deptgpas_coursesGPAavgs}
    \end{axis}
    \end{tikzpicture}
    \begin{tikzpicture}
    \begin{axis}[title=\textbf{DeptGPAsMedian\_couorsedeptGpas},xlabel={Time (min)},ylabel={NextComparison score},ylabel near ticks,
    xmin=0,xmax=240,xtick={0,60,120,...,240},xmode=linear,ymax=200,ymin=0,ytick={50,100,150,200},width=0.25\textwidth]
       \input{plots/time-series-subplots/deptgpas_coursedeptGPas}
    \end{axis}
    \end{tikzpicture}
    \begin{tikzpicture}
    \begin{axis}[title=\textbf{DeptGPAsMedian\_median},xlabel={Time (min)},ylabel={NextComparison score},ylabel near ticks,
    xmin=0,xmax=240,xtick={0,60,120,...,240},xmode=linear,ymax=400,ymin=0,ytick={100,200,...,400},width=0.25\textwidth]
       \input{plots/time-series-subplots/deptgpasmedian_median}
    \end{axis}
    \end{tikzpicture}

%% file: case_studies/case_study_stocks.tex
\newcommand{\StockProgramName}{\textit{StockBuyAndSell}\xspace}
\newcommand{\maxProfitsRDD}{\textit{maxProfits}\xspace}

\subsection{Case Study: StockBuyAndSell}

Suppose a developer is interested in the \StockProgramName program, which is based on the LeetCode problem~\cite{leetcode-buy-and-sell-iii}. Using a dataset of comma-separated strings in the form \textit{"Symbol,Date,Open,High,Low,Close,Volume,OpenInt"}, the \StockProgramName calculates each stock's maximum achievable profit with at most three transactions (using a dynamic programming implementation adapted from \cite{leetcode-buy-and-sell-iii-peetleetcode-solution}) by grouping closing prices by stock symbol and sorting within each symbol. 
As an initial dataset, the developer samples 1\% of the 20 largest stock symbols from a Kaggle dataset \cite{Kaggle-Stock-Data-Marjanovic2017}. The dataset consists of 2,389 records across 20 partitions.

The developer wishes to generate an input that triggers a symptom where one partition increases the maximum observed profit of the dynamic programming loop at least five times more frequently than other partitions. For the target UDF, they specify the RDD variable that is produced from the transformation in which this loop occurs.
The developer may use Spark's Accumulator API~\cite{spark} to count the number of branch executions to model an increase of the maximum observed profit for each partition. This metric is then passed to a \textit{NextComparison} monitor template with a ratio of 5.0.

The target UDF takes \textit{(String, Iterable[Double])} tuples as input. The developer uses their knowledge of the \StockProgramName program to disable key-based mutations for these inputs, as well as impose a restriction that UDF input keys must be unique due to an earlier aggregation. As a result, \perfgen produces the following two mutations: \textsc{M10 + M7 + M5 + M2} and \textsc{M10 + M7 + M6}.

Finally, \perfgen generates a pseudo-inverse function with GPT-4. The generated function first assigns a chronological date to each price within a stock group. Next, it populates arbitrary values for unused program input fields. Lastly, it joins all values into the comma-separated string format required by \StockProgramName. 

\begin{minipage}{\linewidth}
\input{code/code_stockbuysell_inverse_fn}
\end{minipage}

\perfgen starts generating test inputs by first partially executing \StockProgramName on the provided input dataset to produce a UDF input consisting of stock symbols and their chronologically ordered prices.
It then applies mutations to this intermediate input and, after 3.4 mins and 4,775 iterations, produces an input which satisfies the monitor template. The result is produced from a \textit{M5} which directly affects the developer's custom metric by modifying individual values in the grouped stock prices.

Next, \perfgen applies the pseudo-inverse function to the UDF input, tests the resulting \StockProgramName input, and finds that it also triggers the symptom of interest. As a result, no additional fuzzing iterations are necessary.

For comparison, we evaluate the \StockProgramName program using the initially provided input dataset and the baseline configuration discussed at the start of Section \ref{sec:evaluation}. After approximately 4 hours and 40,010 iterations, no inputs trigger the symptom of interest. 

\StockProgramName evaluation results are summarized in Table \ref{table:case_study_results}, with the progress of the best observed \textit{NextComparison} ratios plotted in Figure \ref{fig:time_series}.
Compared to the baseline which times out after four hours, \perfgen leads to at least 69.46X speedup and requires at most 0.002\% of the program fuzzing iterations. Additionally, 98.91\% of \perfgen's execution time is spent on UDF fuzzing alone. 

While \perfgen is able to trigger the performance symptom of interest--a \textit{Next Comparison} ratio greater than 5.0, the baseline only reaches a ratio of approximately 2.5, indicating a substantial gap in the two approaches' effectiveness. 
This is because the baseline is unable to handle fields that are unused or parsed into numbers, nor is it able to significantly affect the distribution of data across each key. In contrast, \perfgen overcomes these challenges through its phased fuzzing and tailored mutations.

%% file: code/code_stockbuysell_inverse_fn.tex
\begin{lstlisting}[language=Scala]
def inverse(udfInput: RDD[(String, Iterable[Double])]): RDD[String] = {
  val datePrice = udfInput.flatMapValues(prices => {
    prices.map(price => {
      val dateStr = getNextDate()(dateStr, price)})})
  val stringJoin = datePrice.map({
  case (key, valueTuple) => val (date, price)=valueTuple
    // "Symbol,Date,Open,High,Low,Close,Volume,OpenInt"
    Seq(key, date, price, price, price, 100000, 0).mkString(",")})
  return stringJoin}
\end{lstlisting}

%% file: case_studies/case_study_dept.tex
\newcommand{\groupedRDD}{\textit{grouped}\xspace}

\subsection{Case Study: DeptGPAsMedian}
\label{sec:case_study_deptgpas}

\begin{figure}[!th]
 \centering 
 \input{code/code_deptgpamedian_program}
 \vspace{-3ex}
 \caption{The \textit{DeptGPAsMedian} program implementation in Scala which calculates the median of average course GPAs within each department.}
 \label{fig:code_dept}
 \vspace{-2ex}
\end{figure}

Consider a scenario in which a developer is examining the \textit{DeptGPAsMedian} program (as depicted in Figure~\ref{fig:code_dept}), which is a modified version from \cite{flowdebug}. This program operates on a string dataset structured in the format \textit{"studentID,courseID,grade"}, and initially computes the average GPA for each course. It then groups these average GPAs based on the department associated with each course and calculates the median average course GPA for each department. In order to scrutinize this program, the developer generates a 40-partition dataset encompassing 5,000 records per partition, thereby amassing a total of 2.8MB data. This dataset comprises five departments, each offering 20 courses, and encompasses 200 distinct students.

In the absence of a specified target stage (as indicated in the preceding three examples), \perfgen targets phase fuzzing at each stage. To better quantify the data skew symptom of interest, the developer seeks to generate a dataset in which a single post-aggregation partition reads a minimum of 100 times the number of shuffle records compared to the other partitions. Leveraging \perfgen, the developer outlines their symptom using the \textit{Shuffle Read Records} metric and a \textit{NextComparison} monitor template configured with a target ratio of 100.0. Anticipating small intermediate partitions (with 100-course averages spread across 40 partitions), they adjust \perfgen to use an elevated duplication factor of 5$\times$, thereby inducing data skew associated with the \textit{Shuffle Read Records} metric. Analogous to the earlier example, \perfgen computes relevant mutations and sampling weights, giving data skew-oriented mutations larger weights and, consequently, increased sampling probabilities.

Finally, \perfgen leverages the large language model GPT-4 to generate pseudo-inverse functions that map each stage back to the original input. This is accomplished by prompting the question, ``Could you produce inverse functions that transition from \codefont{val courseGrades/courseGpaAvgs/deptGpas/median} to the initial input?'' It takes less than 20 seconds and 1 query to get answers for each pseudo-inverse function. The average LoC of pseudo-inverse functions is 6.

For instance, consider the stage \codefont{val courseGrades}. The pseudo-inverse function generated for this stage is as follows:

\begin{minipage}{\linewidth}
\input{code/code_deptgpamedian_inverse_fn}
\end{minipage}
This pseudo-inverse function generates the original input from the intermediate data \textit{"(courseID, average grades)"} to the original input, which is a string  in the format \textit{"studentID,courseID,grade"} For example, applying this pseudo-inverse function on an RDD containing a single intermediate (UDF) record of \textit{("EE", 80.7}) produces an RDD containing a single \DeptProgramName input record of \textit{"42,EE,80"}. Running the \DeptProgramName program with this input produces an intermediate RDD of a single record, \textit{("EE", 80.0)}, which differs from \textit{("EE", 80.7}).  Nonetheless, the output of the pseudo-inverse function (i.e., the generated \DeptProgramName input RDD) is still a valid, improved seed to the \DeptProgramName program.

\perfgen starts generating test inputs by partially executing \DeptProgramName on the provided input data to derive UDF inputs consisting of each course's department name paired with the course's average grade. It then applies mutations to generate new intermediate inputs and tests them to see if they trigger the target data skew symptom. After 4.3 minutes and 1,519 iterations, \perfgen produces an input by using a \textit{M13} mutation which significantly increases the frequency of UDF inputs associated with the "EE" department.
Next, it applies the pseudo-inverse function to this intermediate UDF result to produce a \DeptProgramName input which contains thousands of unique courses in the "EE" department. It then tests this input with the full \DeptProgramName program and finds that the target symptom is triggered with no additional modifications.

For comparison, we evaluate \DeptProgramName using the same initial dataset and the baseline configuration specified earlier at the start of Section \ref{sec:evaluation}. Under these settings, the baseline is unable to generate any symptom-triggering inputs after approximately 4 hours and 21,575 iterations.

The \DeptProgramName case study results are summarized in Table \ref{table:case_study_results}, and the progress of the best observed \textit{NextComparison} ratios are displayed in Figure \ref{fig:time_series}. Using \perfgen over the baseline configuration produces at least 54.80X speedup while requiring at most 0.005\% of the program fuzzing iterations.\perfgen's UDF fuzzing process comprises 98.61\% of its total execution time.

While \perfgen is able to trigger the performance symptom of interest--\textit{Next Comparison} ratio greater than 100.0. On the other hand, the baseline is unable to reach even 7\% of this threshold. This gap in performance can be attributed to baseline's inability to target data skew symptoms. On the other hand, \perfgen can precisely target the appropriate stage in the Spark program through its use of UDF fuzzing, and is able to leverage skew-oriented mutations to modify the underlying data distribution, causing data skew. 

%% file: code/code_deptgpamedian_program.tex
\begin{lstlisting}[language=Scala]
val lines =  HybridRDD(sc.textFile("grades"))

val courseGrades = lines.map(line => {
  val arr = line.split(",")
  val (courseId, grade) = (arr(1), arr(2).toInt)
  (courseId, grade)
})

// assign GPA buckets
val courseGpas = courseGrades.mapValues(grade => {
  if (grade >= 93) 4.0
  else if (grade >= 90) 3.7
  else if (grade >= 87) 3.3
  else if (grade >= 83) 3.0
  else if (grade >= 80) 2.7
  else if (grade >= 77) 2.3
  else if (grade >= 73) 2.0
  else if (grade >= 70) 1.7
  else if (grade >= 67) 1.3
  else if (grade >= 65) 1.0
  else 0.0
})

// Compute average per key
val courseGpaAvgs = 
  courseGpas.aggregateByKey((0.0, 0))(
    { case ((sum, count), next) => 
      (sum + next, count + 1) },
    { case ((sum1, count1), (sum2, count2)) => 
      (sum1 + sum2, count1 + count2) }
  ).mapValues({ case (sum, count) => 
      sum.toDouble / count })

val deptGpas = courseGpaAvgs.map({ case (courseId, gpa) =>
  val dept = courseId.split("\\d", 2)(0).trim()
  (dept, gpa)
})

// Use 3 partitions due to few keys
val grouped = deptGpas.groupByKey(3)

val median = grouped.mapValues(values => {
  val sorted = values.toArray.sorted
  val len = sorted.length
  (sorted(len / 2) + sorted((len - 1) / 2)) / 2.0 })

\end{lstlisting}

%% file: code/code_deptgpamedian_inverse_fn.tex
\begin{lstlisting}[language=Scala]
def inverse(udfInput: RDD[(String, Double)]): RDD[String] = {
    rdd.zipWithUniqueId().map({
        case ((course, avg), uniqueID) =>
        val courseStr = course + uniqueID, grade = avg.toInt
        s"0,$courseStr,$grade"})} // dummy student ID
\end{lstlisting}

%% file: plots/rq3_dept_gpa_time.tex
\begin{tikzpicture}
    \begin{axis}[
        xmin=0,
        height=0.25\textwidth,
        width=0.5\textwidth,
        xtick distance = 10,
        xmode=linear,
        xlabel={Mutation Selection Probability (\%)},
        ylabel={Input Generation Time (minutes)},
        ylabel near ticks,
        xlabel near ticks,
    ]
        \addplot[ultra thin,smooth,mark=*] plot coordinates {

        (2.44, 34.58)
        (11.11, 13.71)
        (20.00, 6.12)
        (38.46, 5.39)
        (55.56, 3.39)
        (65.22, 3.83)
        (71.43, 1.43)

    };
    \end{axis}
\end{tikzpicture}

%% file: related.tex
\section{Related Work}

\noindent{\bf Performance Analysis of Big Data Systems.}
Today, big data analytics applications are always composed of both UDF and dataflow operators. Performance analysis has been a long challenge in big data systems~\cite{yu2018datasize,hanel2020vortex,nguyen2018skyway,Venkataraman:Ernest}. Ernest \cite{Venkataraman:Ernest}, ARIA \cite{Verma:ARIA}, and Jockey \cite{Ferguson:Jockey} model job performance by observing system and job characteristics. Starfish \cite{Herodotou:Starfish} constructs performance models and proposes system configurations that either meet the budget or deadline requirements. DAC~\cite{yu2018datasize} is a data-size aware auto-tuning approach to efficiently identify the high dimensional configuration for a given Apache Spark program to achieve optimal performance on a given cluster. It builds the performance model based on both the size of input dataset and Spark configuration parameters. Cheng et al.~\cite{cheng2021efficient:efficientsparkprediction} incorporate up to 180 Spark configuration parameters to predict Spark application performance for a given application and dataset size. They do so by training Adaboost ensemble learning models to predict performance at each stage, while minimizing required training data through a data mining technique known as projective sampling.

Kwon et al.~present a survey of various sources of performance skew and identify different kinds of data skews~\cite{Kwon:SkewStudy}. Several works aim to mitigate the performance impacts of data skew by proposing automatic skew mitigation approaches that leverage techniques such as data redistribution and adaptive repartitioning~\cite{Kwon:SkewTune,chen2014libra,liu2015dreams,Bindschaedler:Hurricane,Tang:IntermediatPartitionSkewMitigationSpark,Liu:SPPartitioner}. Mishra et al. \cite{Mishra:HadoopSkewnessMethods} conduct a survey of Hadoop-based data skewness mitigation techniques and categorize them based on each technique's support for map-side and reduce-side data skew.

All these performance analysis techniques assume that benchmarks provide insights about workload characteristics and performance depends on data size and cluster configurations. However, many ignore the fact that performance is often also dependent on input content. They do not auto-generate inputs to trigger performance skews. In contrast, \tool automatically generates such performance workload to reproduce a desired performance symptom.

\vspace{.3em} 
\noindent\textbf{Test Generation for DISC Applications.}
State of the art test generation techniques for DISC applications fall into two main categories: symbolic-execution based approaches~\cite{Gulzar2020:BigtestDemo,li2013sedge,Olston2009sigmod} and fuzzing-based approaches~\cite{zhang2020:bigfuzz}. Gulzar et al.~model the semantics of these operators in first-order logical specifications alongside with the symbolic representation of UDFs~\cite{Gulzar2020:BigtestDemo} and generate a test suite to reveal faults. Prior DISC testing approaches either do not model the UDF or only model the specifications of dataflow operators partially~\cite{li2013sedge,Olston2009sigmod}. Li et al.~propose a combinatorial testing approach to bound the scope of possible input combinations~\cite{bittag}. All these symbolic execution approaches generate path constraints up to a given depth and are thus ineffective in generating test inputs that can lead to deep execution and trigger performance skews. To reduce fuzz testing time for dataflow-based big data applications, BigFuzz~\cite{zhang2020:bigfuzz} rewrites dataflow APIS with executable specifications; however, its guidance metric concerns branch coverage only and thus cannot detect performance skews. Additionally, there is no guarantee that the rewritten program preserves the original DISC application's performance behaviors.

\vspace{.3em} 
\noindent\textbf{Fuzz Testing for Performance.} 
Fuzzing has gained popularity in both academia and industry due to its black/grey box approach with a low barrier to entry~\cite{afl}. For example, AFL mutates a seed input to discover previously unseen branch coverage~\cite{afl}. Instead of using fuzzing for code coverage, several techniques have investigated how to adapt fuzzing for performance testing. PMFuzz~\cite{liu2021pmfuzz} generates test cases to test the crash consistency guarantee of programs designed for persistent memory systems. It monitors the statistics of PM paths that consist of program statements with PM operations. PerfFuzz~\cite{lemieux2018perffuzz} uses the execution counts of exercised instructions as fuzzing guidance to explore pathological performance behavior. MemLock~\cite{memlock:2020:fuzzing} employs both coverage and memory consumption metrics to guide fuzzing; however, it does not leverage a phased fuzzing approach to expedite fuzzing time and is limited to uncontrolled memory consumption bugs only.

Unlike these works, \tool tackles the performance testing challenge of DISC applications. It leverages a phased fuzzing approach to expedite the test generation process, uses performance skew symptoms as a fuzzing objective, and generates inputs to replicate these symptoms through custom performance monitoring and skew-inspired mutations. 

\vspace{.3em}
\noindent\textbf{Program Synthesis for Data Transformation.}
Inductive program synthesis~\cite{Gulwani2010:programsynthesis} learns a {\em program} (i.e., a procedure) from incomplete specifications such as input and output examples. FlashProfile~\cite{padhi2017flashprofile} adapts this approach to the data domain, presents a novel domain-specific language (DSL) for patterns, defines a specification over a given set of strings, and learns a syntactic pattern automatically. PADS~\cite{Fisher2011:PPO} provides a data description language allowing users to describe their ad-hoc data for various fields in the data and their corresponding type. The data description is then generated automatically by an inference algorithm. Oncina et al.~\cite{Oncina92identifyingregular} propose a new algorithm that learns a DFA compatible with a given sample of positive and negative examples.  However, a key limitation of prior work is that such data transformation synthesis is not used to speed up fuzz testing. \perfgen addresses this gap and is the first to employ OpenAI's ChatGPT~\cite{openai2023gpt4} to generate a pseudo inverse function  to produce improved seeds to be used in the beginning of the program and thus reduce the overall fuzz testing time. This approach not only integrates program synthesis with test input generation but also utilizes the power of advanced large language models, providing a more robust and effective solution for data transformation tasks.

%% file: conclusion.tex
\section{Conclusion}

This paper presents \tool, a tool that automatically generates inputs for reproducing a desired performance symptom in a DISC application. \tool's novel phased fuzzing approach addresses the challenge of latency reduction, symptom-specific performance feedback monitoring, and skew-inspired mutations. \tool marks the first instance of a tool leveraging a large language model to enhance performance testing, achieved by generating pseudo-inverse functions. Our evaluation shows that \tool achieves an average speedup of at least \EvalResultSpeedupX compared to traditional fuzzing approaches when generating workloads that exhibit specified performance skew symptoms. Furthermore, it does so while requiring at most \EvalResultIterPct of the fuzzing iterations when phased fuzzing is disabled. Finally, our evaluation explores the impact of mutation sampling probabilities on input generation efficiency and finds that \tool's template-inspired mutation probabilities produce a \EvalResultMutationSpeedup speedup in input generation time compared to a uniform sampling approach.